\documentclass{aa}  

\newcommand{\Herschel}{{\it Herschel }}

\usepackage{graphicx}
\usepackage{txfonts}

\begin{document} 

\title{Comparison of modified black-body fits for the estimation of dust optical
depths in interstellar clouds}

\author{M. Juvela}
\institute{
Department of Physics, P.O.Box 64, FI-00014, University of Helsinki,
Finland, {\email mika.juvela@helsinki.fi}
}

\date{Received September 15, 1996; accepted March 16, 1997}

\abstract
{When dust far-infrared spectral energy distributions (SEDs) are fitted with a
single modified black body (MBB), the optical depths tend to be underestimated.
This is caused by temperature variations, and fits with several temperature
components could lead to smaller errors.}
{We want to quantify the performance of the standard model of a single MBB in
comparison with some multi-component models. We are interested in both the
accuracy and computational cost.}
{We examine some cloud models relevant for interstellar medium studies.
Synthetic spectra are fitted with a single MBB, a sum of several MBBs, and a
sum of fixed spectral templates, but keeping the dust opacity spectral index fixed.}
{When observations are used at their native resolution, the beam
convolution becomes part of the fitting procedure. This increases the computational
cost, but the analysis of large maps is still feasible with direct
optimisation or even with Markov chain Monte Carlo methods. Compared to the
single MBB fits, multi-component models can show significantly smaller
systematic errors, at the cost of more statistical noise. The $\chi^2$
values of the fits are not a good indicator of the accuracy of the $\tau$
estimates, due to the potentially dominant role of the model errors. The
single-MBB model also remains a valid alternative if combined with empirical
corrections to reduce its bias.}
{It is technically feasible to fit multi-component models to maps of
millions of pixels. However, the SED model and the priors need to be selected
carefully, and the model errors can only be estimated by comparing alternative
models.}


\maketitle

\section{Introduction} \label{sect:introduction}

Thermal dust emission at far-infrared and millimetre wavelengths is used to
trace the structure and physical conditions of interstellar matter (ISM),
especially in connection with star formation (SF). The targets range from
a Galactic diffuse medium and local molecular clouds to entire galaxies. The
measured intensities represent emission from a range of conditions that depends on
both the nature of the sources and the angular resolution of the
observations. The continuum emission also carries information about the dust
itself, which is interesting due to the significant and systematic evolution
that takes place in the dust properties during the SF process
\citep{Draine2003ARAA}.

The long-wavelength dust emission is often modelled as modified black-body
(MBB) emission, a black-body spectrum multiplied by a power law representing
the frequency dependence of the dust absorption cross section $\kappa_{\nu}$.
The analysis may assume both a single temperature $T$ and optically thin
emission, which is only an approximation of the emission of any astronomical
source. If the assumptions of a single temperature and optically thin
emission are both violated, an analytical estimation of source properties is
hardly possible, and more detailed modelling of the observations is needed.
Even for optically thin sources, all observations consist of emission from
dust at different temperatures within the telescope beam, both along the line
of sight (LOS) and over the sky. The frequency dependence of $\kappa_{\nu}$
is also more complex than a single power law \citep{Paradis2010,
planck2013-p06b, planck2013-XVII, Gordon2014, Kohler2015, GCC-VI}. However,
given the limitations as to the number of observed frequencies and the
signal-to-noise ratio (S/N) of the data, the spectral energy distributions
(SEDs) have to be modelled with just a few free parameters, which include at
least the intensity and one temperature. The simultaneous fitting of
temperature and $\beta$ is complicated by the inherent anti-correlation of
these parameters, which makes the individual parameters very sensitive to
noise \citep{Shetty2009b, Kelly2012,
Juvela2012_bananasplit,Juvela2013_TBmethods}. The problem is further
complicated by temperature variations, which tend to systematically decrease
the observed $\beta$ values relative to the actual $\beta$ of the dust
\citep{Shetty2009a,Juvela2012_Tmix}. If dust emission is only used to measure
column densities, the results are likely to be more robust if $\beta$ is kept
constant, even if the real variations of $\beta$ may then result in up to
tens of percent errors in the column density estimates. Spectral indices can
only be constrained with high-S/N data that cover a wide frequency range, and
even then the estimates can still be biased due to the temperature
variations.

The single-component MBB model can be extended by allowing multiple
temperature or $\beta$ components. In studies of dust properties, one may
further allow changes in $\beta$ as a function of frequency
\citep{Paradis2012_500um, Gordon2014, Galliano2018}. However, many sources
are known to have wide temperature distributions, with smaller effects
expected from the $\beta$ variations. Therefore, it is more common to use
multiple temperature components while the dust optical properties are assumed
to be constant or modelled with one free parameter at most. This is the case
especially in studies of external galaxies, where most temperature structures
remain unresolved \citep{Galametz2012, Chang2021}. However, temperature
variations can be significant even in local SF clouds, from $\sim$6\,K inside
cold cores \citep{Crapsi2007,Harju2008,Pagani2015} to more than 100\,K in hot
cores \citep{Motte2018,Jorgensen2020}. In Galactic studies, the
single-component MBB model is still the main tool, although it suffers from
the well-known tendency to underestimate column densities. The actual
three-dimensional temperature fields can only be estimated with radiative
transfer modelling or, in the case of geometrically simple objects, with
tools such as the inverse Abel transform \citep{Roy2014}.

The calculation of multi-component MBB fits poses some technical challenges.
The increased number of free parameters may require regularisation, for
example by using parameter priors in the Bayesian framework. Especially the
problem of $T-\beta$ correlations has led to the use of hierarchical models,
where the hyper-parameters of the parameter distributions are fitted together
with the parameters (i.e. $T$ and $\beta$) of individual measurements
\citep{Kelly2012,Galliano2018,Lamperti2019}. The efficiency of priors and
hierarchical models still varies from case to case. In a complex SF
region, the priors must accommodate a wide range of temperatures and provide
correspondingly weaker constraints for the individual measurements. There is
also the risk of biasing the estimates for rare or unexpected sources, such
as an individual hot (or cold) core seen against a field of colder (or
warmer) background \citep{Juvela2013_TBmethods}.

\citet{Marsh2015} have presented the point process mapping ({\sc PPMAP}) method,
which fits dust observations with a combination of fixed SED templates, such
as MBB functions \citep[e.g.]{Marsh2017,Howard2019}. The emission is
represented by points in the state phase, and a requirement of using the
minimum set of points to fit the data (to an accuracy of $\chi^2\sim1$) also
acts as regularisation. One important point is that {\sc PPMAP} uses all
observations at their native resolution. The more typical approach is to
convolve all observations to a common lower resolution, which then allows the
SED of each map pixel to be analysed independently. The use of the native data
resolution brings additional challenges to the SED fitting. First, it requires
higher S/N observations since the data are used at a higher resolution (more
noise per resolution element) and the short-wavelength information of
small-scale intensity variations is combined with weaker constraints on the
SED shape, which partly depend on observations at longer wavelengths and
lower angular resolutions. Second, the fitting can no longer be done pixel by
pixel; it has become a global optimisation problem with a much higher
computational cost.

In this paper, we use simulations of interstellar clouds to compare the
precision of column density estimates calculated with the single-component MBB
model and with some multi-component models. For the latter, we examine both
the use of SED templates (here MBB functions with fixed temperatures) and the
use of MBB functions with temperatures directly as free parameters. The
observations are used at their original resolution by including the
convolution with the telescope beam as part of the model. In view of the
associated increased computational cost, we also pay attention to the
organisation of calculations and the potential benefits of using graphics
processing units (GPUs). The fits were mainly carried out with direct numerical
optimisation, but some tests were also carried out using Markov chain Monte
Carlo (MCMC) methods. The simulated observations include simple toy models
with two temperature components and more complex cloud models where the
synthetic observations were created through radiative transfer modelling.

The contents of the paper are the following. In Sect.~\ref{sect:methods} we
present the multi-component MBB models and discuss some details of the SED
fits. Section~\ref{sect:results} presents the main results from the tests,
which include toy models with ad hoc temperature distributions
(Sect.~\ref{results:toy}), Bonnor-Ebert spheres with internal and external
heating (Sect.~\ref{results:BE}), a cloud filament with stochastically heated
grains (Sect.~\ref{sect:filament}), and a large-scale cloud model based on
magnetohydrodynamic (MHD) simulations (Sect.~\ref{results:IRDC}). We discuss
the results in Sect.~\ref{sect:discussion} before presenting the main
conclusions in Sect.~\ref{sect:conclusions}.

\section{Methods} \label{sect:methods}

Dust emission can be approximated as MBB emission, where the
observed intensity $I_{\nu}$ is
\begin{equation}
I_{\nu} = B_{\nu}(T, \beta) (1-e^{-\tau_{\nu}}) \approx B_{\nu}(T) \tau_{\nu} =
B_{\nu}(T) \Sigma \kappa_{\nu}(\nu_0) \left(  \frac{\nu}{\nu_0} \right)^{\beta},
\label{eq:MBB0}
\end{equation}
where $T$ is in principle the dust temperature but more precisely the colour
temperature that depends on all temperatures along the LOS. In
Eq.~(\ref{eq:MBB0}) the optical depth $\tau$ is written as the product of
mass surface density $\Sigma$ and the dust mass absorption coefficient
$\kappa_{\nu}$, which is further assumed to follow a power law with the
opacity spectral index $\beta$. The constant $\nu_0$ is an arbitrary
reference frequency. In the following, Eq.~(\ref{eq:MBB0}), the modified
black-body function, is written as $MBB(T,\beta)$.

When the emission originates from a source with a distribution of
temperatures, the intensities may still be approximated with a sum of MBB
function,
\begin{equation}
I_{\nu}  \approx \sum_i  W(T_i) MBB(T_i, \beta_i),
\label{eq:MBB}
\end{equation}
where the weights $W(T_i)$ are the fitted parameters. We use two versions of
Eq.(~\ref{eq:MBB}). In the $N$-MBB models, the fit is based on directly on
the above equation at each map position $\bar{r}$,
\begin{equation}
I(\bar{r},{\nu}) = \sum_i^{N} W_i(\bar{r}) \,\, MBB_{\nu}(T_i(\bar{r}), \beta_i({\bar r})).
\label{eq:a}
\end{equation}
The model includes $N$ MBB functions, with $W_i(\bar{r})$ and $T_i(\bar{r})$
as the free parameters while the spectral index $\beta_i(\bar{r})$ is kept
constant. With $N$=1, Eq.(\ref{eq:a}) describes the standard single-component
MBB fit (1-MBB).

The alternative SED models $N$-TMPL correspond to
\begin{equation}
I_{\nu} = \sum_i^{N} W_i (\bar{r})  \,\, C_i(\nu).
\label{eq:b}
\end{equation}
Here $C_i(\nu)$ are fixed spectral templates. These can be MBB functions for
some fixed values of $T$ and $\beta$, the corresponding functions for some
distributions of $T$ and $\beta$, or any other templates that could be
expected to describe the observed emission. In the tests, we set $C_i(\nu)$
equal to MBB function at a given discrete temperature. Because the
temperatures remain fixed during the fitting, the model is different from
Eq.~(\ref{eq:a}) and the scaling factors $W_i$ are the only free parameters
of the $N$-TMPL models.

In Eqs.(\ref{eq:a})-(\ref{eq:b}), the left side is the observed intensity
$I_{\nu}$, which is the true sky surface brightness convolved with the
telescope beam. In 1-MBB models, we follow the standard analysis, where all
observations are convolved to a common lower angular resolution, the fit
itself does not include convolutions, and each map pixel can be fitted
completely independently. This analysis thus results in estimated intensity
and temperature maps at the same common resolution. In all other cases, the
model and the observations are compared at the original resolution of the
observations. This means that the model predictions have to be convolved with
the telescope beam on each step of fit. The resolution of the resulting
column density maps is now not as well defined. The band with the highest
resolution provides information of the sky brightness at that frequency and
that resolution, while at least two bands are needed to trace temperature
variations. The sensitivity to emission from dust at different temperatures
varies between the bands, making the effective resolution of the resulting
$\tau$ map less precise.

When convolution is part of the optimisation, the fit minimises the weighted
least squares errors
\begin{equation}
\chi^2 = \sum_k  \left( \frac{ I_k   -  M_k \otimes B_k  }{\delta I_k} \right)^2,
\end{equation}
where $k$ refers to the frequency band, $M_k$ is the unconvolved model
prediction (right-hand side of Eq.(\ref{eq:a}) or Eq.(\ref{eq:b})), $B_k$ is
the telescope beam, and `$\otimes$' stands for convolution. The problem is
computationally harder, since the model-predicted surface brightness maps are
convolved with the telescope beams on every step of the optimisation and the
fit parameters for the whole map are connected via the convolutions. The
convolution can be done in real space, with direct summation over the beam
footprint, or using Fourier transformation (fast Fourier transforms, FFTs).
We use mainly the latter option, which is faster, but more limited in its
ability to weight individual measurements or to handle missing values and map
boundaries. 

When beam convolutions are part of the fitting procedure, all parameters for
the entire map are combined into a single optimisation problem. The large
number of data points precludes the use second order information (i.e.
Hessian matrices), which would lead to excessive memory usage. We use mainly
conjugate gradient algorithms, where only the gradient of the $\chi^2$
function is needed (relative to each of the fitted parameters). The problem
remains feasible, because the telescope beams couple data directly only over
small distances. Therefore, although all fitted parameters are in principle
connected, this connection is strong only over short distances.

If we denote the observed surface brightness values with $S$, the $\chi^2$
function of $N$-TMPL models becomes
\begin{equation}
\chi^2 =  \sum_{i,k} \left(  \frac{S_{i,k} - [\sum_c W_{c} C_{c,k}  \otimes B_k]_{i,k}}
{\delta S_{i,k}}  \right)^2,
\end{equation}
where the indices $i$, $k$, and $c$ refer to the position, the frequency
band, and the SED template, respectively. Since the $T_{\rm C}$ values are
pre-selected constants, the number of free parameters ($W_c$) is the same as
the number of the fitted SED components.

For the $N$-MBB models the minimised function is
\begin{equation}
\chi^2 =  \sum_{i,k} \left(  
  \frac{S_{i,k} - \left[ \left(\sum_c M(I_c, T_c, \beta_c \right)  \otimes B_k \right]_{i,k} }
       {\delta S_{i,k}}.
\right)^2.
\end{equation}
Here $M$ stands for the MBB function, and the term in square brackets denotes
the convolved model-predicted map at one frequency. The index $i$ is omitted
within the square brackets, but the result of the convolution is a function
of position (index $i$) and frequency (index $k$). When $\beta$ is kept
constant, the number of free parameters is two times the number of SED
components.
For better performance and to guarantee sufficient accuracy of the gradient
estimates, it is important that the gradients are calculated analytically
(Appendix~\ref{app:MBB}).

By default, the optimisation of $N$-TMPL models is performed in real space.
However, the templates $C_{c,f}$ appear in Eq.~\ref{eq:b} only as constant
factors, and this makes it possible to move the optimisation to Fourier space.
If the Fourier transformation is denoted by $\cal F$, one minimises
\begin{equation}
\chi^2 = \sum_{i^\prime,k} \frac{\left[ {\cal F}(S_{i^\prime,k}) - \sum_c C_{c,k} {\cal
F}(W_{i^\prime,c}) \right]_{i^\prime,k}^2}{dS_k^2}, \label{eq:FOU}
\end{equation}
where the index $i^{\prime}$ now denotes an element in the Fourier space. The
free parameters are the real and imaginary parts of the Fourier amplitudes
that correspond to the real-space weight maps $W_{i,c}$. Because the elements
${\cal F}(W_{i^\prime, c})$ result from the Fourier transformation of a
real-valued function, only $n+(n-1) N$ non-redundant elements need to be
optimised, where $N$ is the map size and $n=\lceil N/2 \rceil$. Within the
optimisation, the costly convolution is replaced with a direct
multiplication. For the largest test cases, this amounts to a speedup of one
or two orders of magnitude. If the observational noise in $S$ is Gaussian,
the errors in the Fourier amplitudes are also normally distributed,
justifying the use of the least-squares formula. One important restriction in
Eq.~(\ref{eq:FOU}) is that the error estimates $dS_{k}$ are now constant
values for each frequency, not being able to be varied pixel by pixel.
Compared to the alternative hypothesis of constant relative uncertainties
over the maps, this leads to differences when a single beam contains widely
different intensity values (changing locally the relative weight of the
different pixels) or when the shape of the observed spectra changes
significantly (changing locally the relative weighting of the different
bands). It is also more difficult to implement general priors or even to
constraint optical depths to positive values (i.e. to enforce $W_{i,k}\ge0$).
In the following, the fits performed in Fourier space are referred to as
$N$-TMPL-F fits.

\section{Results} \label{sect:results}

We compared the performance of the SED models for different simulated
observations. Results are presented for simple toy models
(Sect.~\ref{results:toy}), a Bonnor-Ebert sphere, (Sect.~\ref{results:BE}), a
filament that includes the emission from stochastically heated grains
(Sect.~\ref{sect:filament}), and finally for a more realistic large-scale
magnetohydrodynamic (MHD) simulation (Sect.~\ref{results:IRDC}).

\subsection{Toy models}  \label{results:toy}

\subsubsection{Unconstrained fits}

\begin{figure}
\centering
\includegraphics[width=9cm]{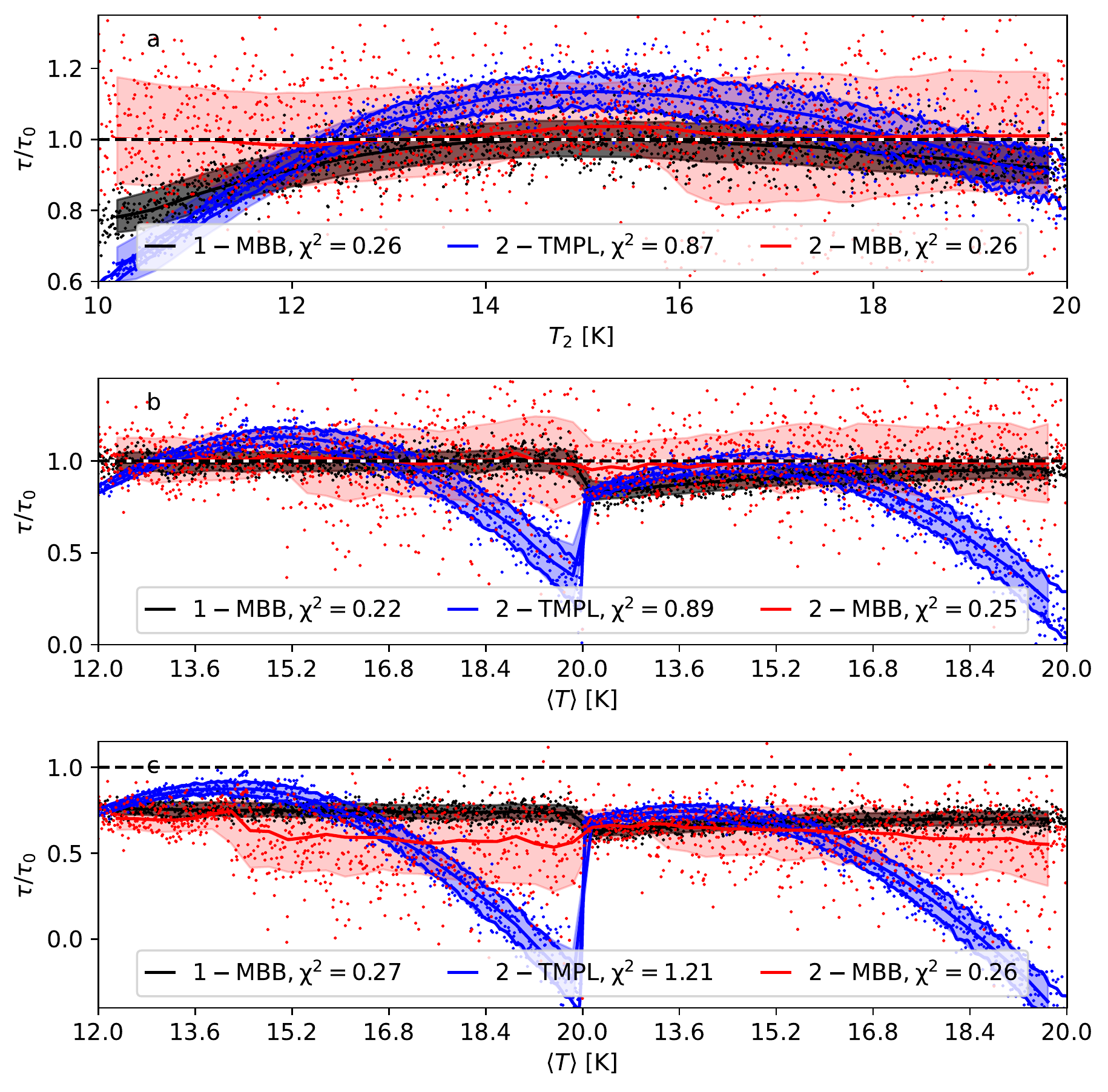}
\caption{
Ratio of estimated and true optical depths $\tau/\tau_0$ for 1-MBB, 2-TMPL
($T_{\rm C}$=13\,K and 17\,K), and 2-MBB models. The observations consist of
160, 250, 350, and 500\,$\mu$m intensities with 3\% relative noise. The dots
correspond to individual pixels (every 10th pixel), the solid lines show
moving averages, and the shaded regions the 1-$\sigma$ dispersion of the
$\tau$ estimates. The $\chi^2$ values of the ML fits (normalised by the
number of fitted data points) are quoted in the frames. In frame a,
observations consist of one emission component at $T_1=15$\,K and one
component varied in the range $T_2$=10-20\,K (on x-axis), both with the same
optical depth.  The additional solid blue lines correspond to the
1-$\sigma$ error estimates from the MCMC fit of the 2-TMPL model. 
In frame b, the observations correspond to a continuous temperature
distribution, $T\sim N( \langle T \rangle, \sigma_T)$. The x-axis shows the
$\langle T \rangle$ value, with $\sigma_T=1$\,K for the first and $\sigma_T=5$\,K for the
second half.
Frame c is the same as frame b but the SED fits assuming $\beta=1.7$ instead
of the correct value of $\beta=1.9$.
}
\label{fig:Figure1}
\end{figure}

To get basic understanding of the performance of the $N$-TMPL and $N$-MBB
models, we started with simple toy models. In Fig.~\ref{fig:Figure1}a, the
observations consist of two emission components, one at a $T_1$=15\,K and
another at a temperature that changes linearly over the 128$\times$128 pixel
maps from $T_2$=10\,K to $T_2$=20\,K. The data include 160, 250, 350, and
500\,$\mu$m maps with 3\% relative observational noise. The beam size was set
in all bands equal to half of the pixel size, to eliminate in this test the
effect of the beam convolutions. This enables more direct comparison between
the 1-MBB and multi-component models, when only the latter have the beam
convolutions as part of the fitted model. The 2-TMPL model uses MBBs at
$T=13$\,K and $T=17$\,K as the SED templates. The data were initially analysed
assuming $\beta=1.9$, the same value that was used to generate the synthetic
observations. Figure~\ref{fig:Figure1} plots the quantity $\tau/\tau_0$, the
ratio between the estimated and the true optical depths.

The 1-MBB model behaves in Fig.~\ref{fig:Figure1}a as expected. The correct
optical depth $\tau$ is recovered when the emission consists of a single
temperature ($T_2 \sim T_1 =15$\,K), and $\tau$ is increasingly
underestimated as the difference between the two temperature components
increases. Since the model has only two free parameters, the scatter of the
optical depth estimates is small.

The 2-TMPL model also has two free parameters ($W_1$ and $W_2$) and the
$\tau$ scatter is similar to that of the 1-MBB model. However, the errors
increase rapidly away from the two selected $T_{\rm C}$ values and are
positive between the two $T_{\rm C}$ values and negative outside the $T_{\rm
C}$ range.

The construction of the simulated data in Fig.~\ref{fig:Figure1}a matches the
assumptions of the 2-MBB model, which also provides practically unbiased
results. However, the number of free parameters is four, and the statistical
noise of the $\tau$ estimates is two times larger than in the 1-MBB and
2-TMPL cases. The absolute accuracy of the individual $\tau$ estimates is
therefore on average not better than for the 1-MBB model. 

Figure~\ref{fig:Figure1}a quotes the $\chi^2$ values of the fits. 
In Fig.~\ref{fig:Figure1} and later, we always quote the $\chi^2$ values
normalised by the number of fitted data points.

The values are not scaled with the degrees of freedom (which is not well
defined for non-linear models or models containing constraints
\citep{Andrae2010a}) and thus directly compares how well the models match the
observed intensities. A smaller $\chi^2$ value (a good match to intensities)
does not necessarily mean more accurate $\tau$ estimates. The 2-TMPL fit
was done with MCMC calculations, and the plotted values are averages over the
MCMC samples. We also plot the 1-$\sigma$ confidence region (containing 68\%
of the MCMC samples), which follows very closely the actual dispersion of the
$\tau$ estimates. However, these do not reflect the true error, which is
dominated by systematic errors.

Figure~\ref{fig:Figure1}b uses a different set of synthetic observations. 
The optical depths are distributed in each pixel at different temperatures
according to a Gaussian $N(T_0, \sigma_T)$. In the plot, $\sigma_T$ is 1\,K
for the first half of the x-axis and 5\,K for the second half. The mean
temperature $\left< T \right>$ increases in both halves linearly from 12\,K
to 20\,K. The temperature distributions are also truncated to $T>8$\,K. The
1-MBB model performs well, with 20\% bias reached only at the lowest
temperatures and with the wider $\sigma_T$5\,K temperature distribution. Most
1-MBB estimates remain within the 1-$\sigma$ error band of the 2-MBB model
that has larger statistical noise. Like in Fig.~\ref{fig:Figure1}a, the
systematic error of the 2-TMPL model are significant. This is true especially
at the highest temperatures, where the dust temperature distribution extends
clearly above the $T_{\rm C}$ values.

Figure~\ref{fig:Figure1}c uses the same observations as frame b but the fits
assume a lower value of $\beta=1.7$, which thus leads to lower $\tau$
estimates. The drop affects the 2-MBB fit slightly more than the 1-MBB fit,
making the latter the most accurate in terms of the $\tau$ estimates. The
wrong $\beta$ value also exacerbates the 2-TMPL errors at high temperatures,
where the estimates now fall even below zero. The unphysical values are the
result of the model trying to match the SEDs (with temperatures outside its
$T_{\rm C}$ range) using a negative weight for the colder component and a
larger positive weight for the warmer component. It is thus clear than the
over 10\,K range of dust temperatures cannot be satisfactorily fitted with
just two $T_{\rm C}$ components.

\subsubsection{Fits constrained with priors}

\begin{figure}
\centering
\includegraphics[width=9cm]{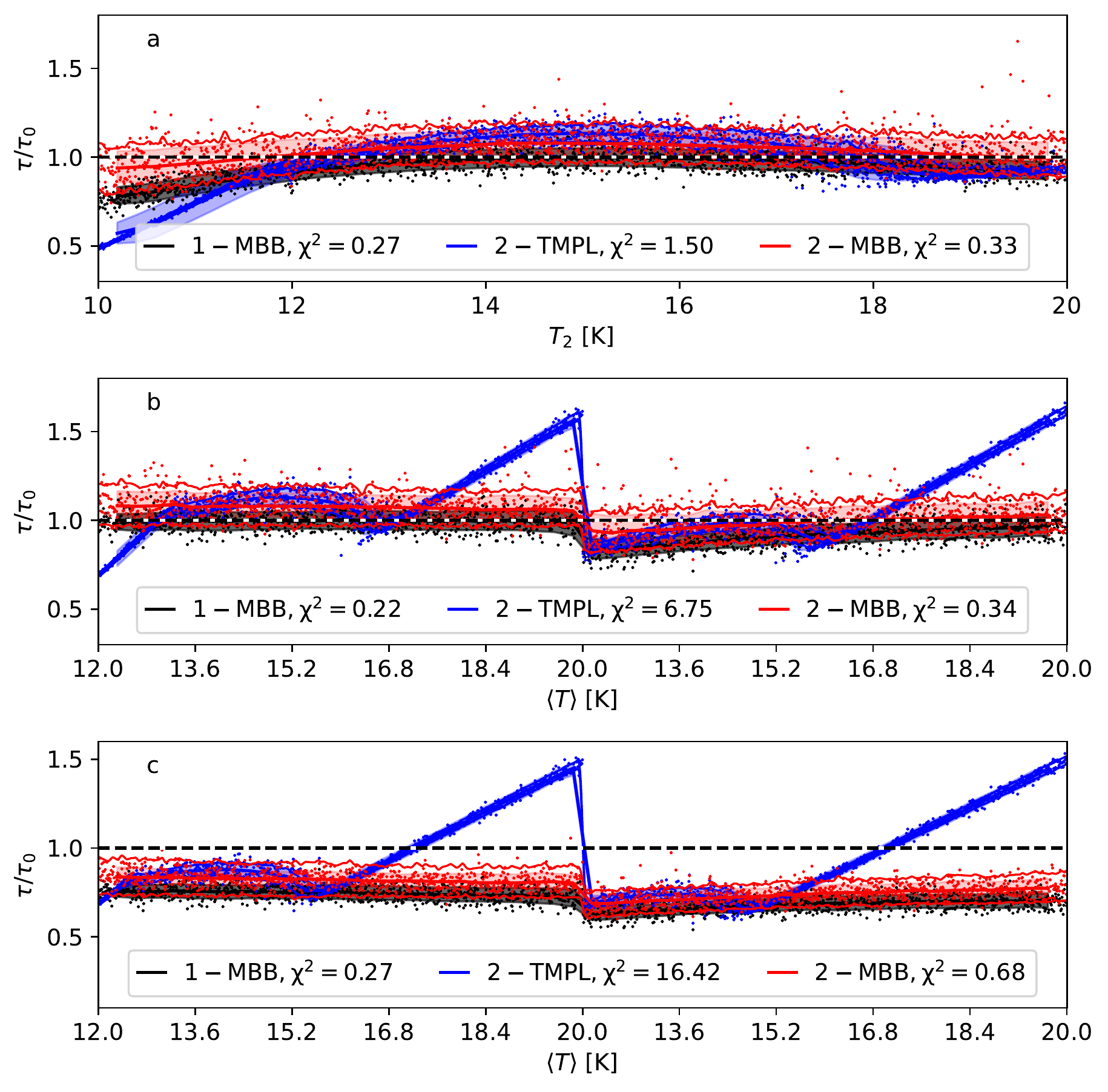}
\caption{
Same as Fig.~\ref{fig:Figure1}, but forcing each component of the 2-MBB and 2-TMPL
models to positive optical depths. In the case of 2-MBB, the model also
includes a prior on the temperatures, $T\sim N(18\,{\rm K}, \/\/ 5\,{\rm K})$.
The plotted 2-TMPL and 2-MBB data are from MCMC calculations while the
$\chi^2$ values are the average values of the maximum-likelihood solutions.
}
\label{fig:Figure2}
\end{figure}

Figure~\ref{fig:Figure2} shows the corresponding results, where we
include a prior $T\sim N(18\,{\rm K}, \/5\,{\rm K})$ for the 2-MBB fits. The
intensities of the 2-MBB fits (parameters $I_{i,c}$) and the 2-TMPL fits
(parameters $W_{i,c}$) are restricted to non-negative values by optimising
the logarithm of the relevant variables (cf. Appendix~\ref{app:MBB}). With
these constraints, both models could be successfully fitted using MCMC.  
There is little difference between the 1-MBB and 2-MBB fits, although the
latter has a smaller bias in the case of the $\sigma(T)=5$\,K observations.
The non-negativity has a clear effect on the 2-TMPL fits, as the optical
depths are now overestimated rather than underestimated for temperatures
above the $T_{\rm C}$ range. This shows that the Fig.~\ref{fig:Figure1}b fit
already contained negative components, even when the total $\tau$ estimates
were positive. Although the unphysical negative values have been eliminated,
the $\chi^2$ values of the fits are now much larger and the systematic errors
of the $\tau$ values still increase rapidly outside the adopted $T_{\rm C}$
range.

\begin{figure}
\centering
\includegraphics[width=9cm]{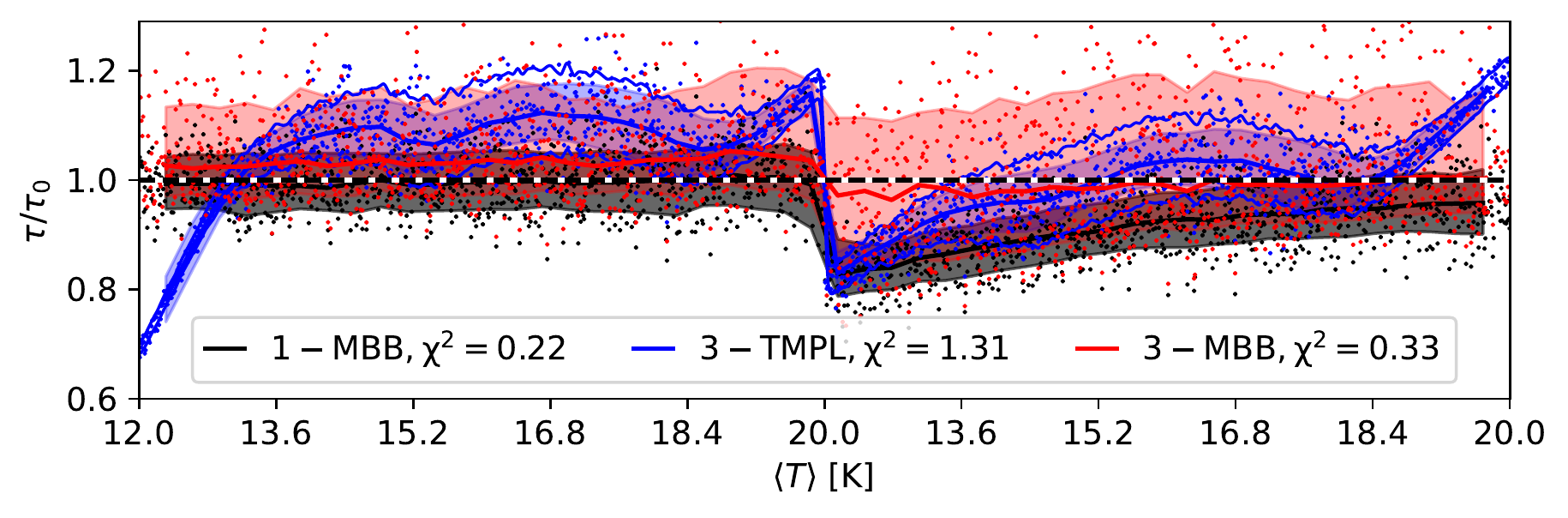}
\caption{
Same as Fig.~\ref{fig:Figure2}b, but including results for 3-TMPL (MCMC fit)
and 3-MBB (ML fit) models. The 3-TMPL model uses $T_{\rm C}=$13, 15.5, and
19\,K, and the 3-MBB model includes a temperature prior $T\sim N(15\,{\rm K},
\/\/ 3\,{\rm K})$.
}
\label{fig:Figure3}
\end{figure}

Figure~\ref{fig:Figure3} shows fits for 3-TMPL and 3-MBB models. With $T_{\rm
C}=$13, 15.5, and 19\,K, the bias in 3-TMPL results can be constrained to
$\sim$20\%. With a prior of $T\sim N(15\,{\rm K}, \/ 3\,{\rm K}$, the results
of 3-MBB fits are unbiased in the $\sigma_{\rm T}=1\,{\rm K}$ part of the map
(left half of the plot) but still underestimate the $\sigma(T)=5$\,K
observations by $\sim$10\%. Even if the systematic errors are partly smaller
than in the 1-MBB fit, there is no clear net improvement because of the
larger statistical noise.

\subsubsection{Fits including beam convolution}

\begin{figure}
\centering
\includegraphics[width=9cm]{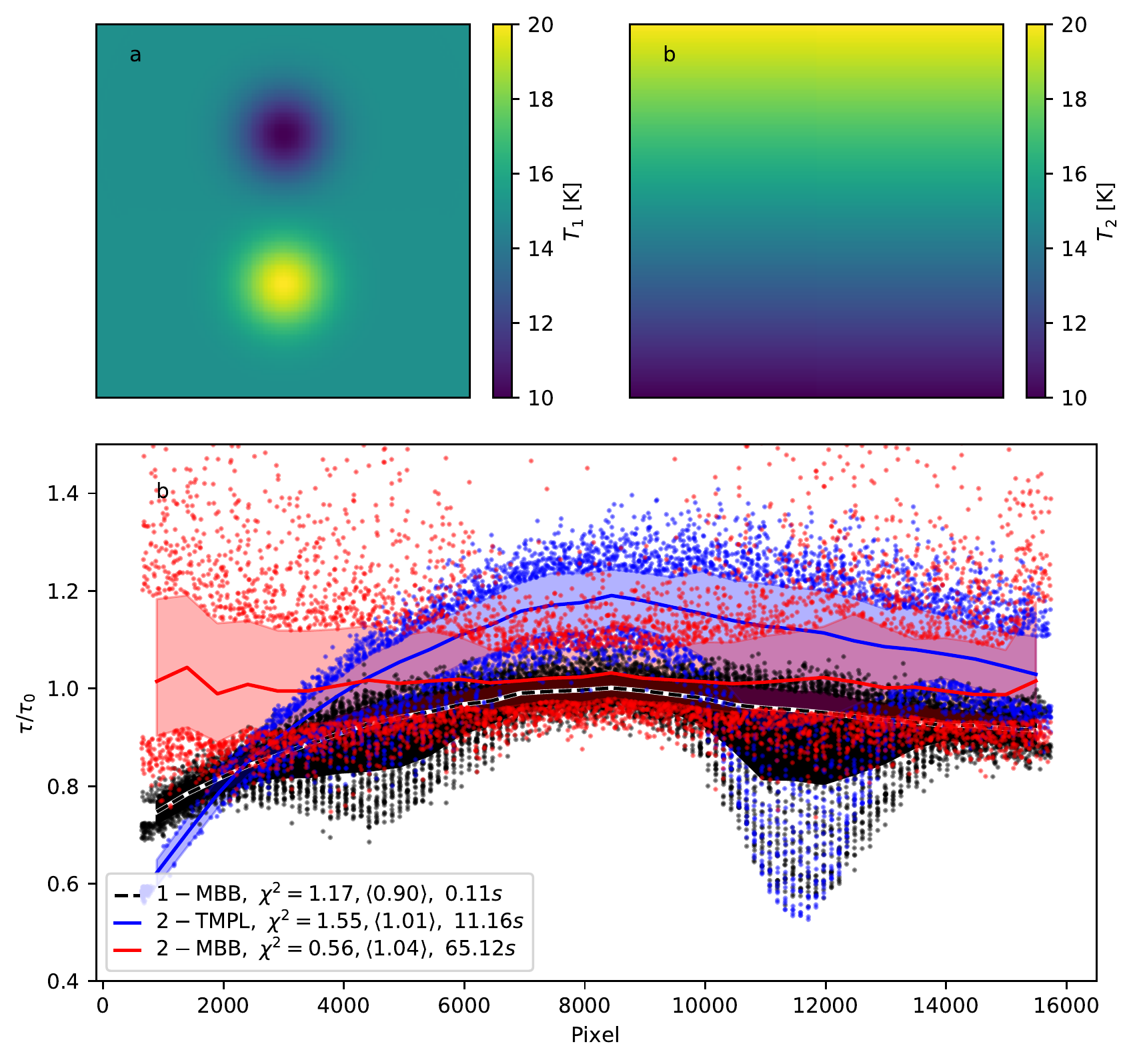}
\caption{
SED fits considering the effect of telescope beams. The 2-TMPL and 2-MBB
cases include beam convolution as part of the fit while the 1-MBB fit is done
at the lowest common resolution. The dust emission consists of the two
temperature components shown in frames a and b.  The map size is
128$\times$128 pixels, the pixel size 10$\arcsec$, and FWHM beam sizes
20$\arcsec$, 20$\arcsec$, 25$\arcsec$, and 36$\arcsec$ at 160, 250, 350, and
500\,$\mu$m, respectively.
}
\label{fig:Figure4}
\end{figure}

Figure~\ref{fig:Figure4} shows the first comparison where the beam size is
different for different bands. The synthetic observations correspond to the
sum of two temperature components. The first one is at 15\,K, apart from two
Gaussian sources where the temperature changes to 10\,K or 20\,K. The second
temperature component increases linearly from 10\,K to 20\,K over the map. The
pixel size is 10$\arcsec$, and the full width at half maximum (FWHM) values of
the Gaussian beams 20, 20, 25, and 36$\arcsec$ at 160, 250, 350, and
500\,$\mu$m, respectively. The observations contain 3\% relative noise at the
resolution of the each map, and all results correspond to the ML solutions.

In Fig.~\ref{fig:Figure4}, the 1-MBB model has again the lowest statistical
errors, partly because all data have been convolved to a $36\arcsec$
resolution. The systematic errors exceed 20\% when the 15\,K component
is combined with the extended second component at $\sim$10\,K. (leftmost part
of Fig.~\ref{fig:Figure4}c), when the warm Gaussian source is combined with
cold extended emission (around pixels index $\sim$4500), and even more
significantly when the warm 17-18\,K extended emission is combined with the
cold Gaussian source (pixel indices $\sim$12000). 

The 2-TMPL model uses values $T_{\rm C}$=13\,K and 18\,K. The statistical
noise is larger than for the 1-MBB model but now also corresponds to a factor
of two higher angular resolution. The angular resolution of the model is
nominally $20\arcsec$, but because it uses all bands up to the 500\,$\mu$m map
with 36$\arcsec$ resolution, the solution might exhibit weak correlations up
to that scale. The $\tau$ values are again significantly underestimated, when
the emission contains emission from dust below the $T_{\rm C}$ values. At the
location of the cold Gaussian source (10\,K against the $\sim$17-18\,K
background), the bias reaches 50\%, the same as for the 1-MBB model. Towards
the map centre, the emission corresponds to the single 15\,K temperature and
$\tau$ is overestimated by up to 20\%.

The results of the 2-MBB model are practically unbiased. However, apart from
the effect of convolutions, the synthetic observations also precisely match
the assumptions of the 2-MBB model. The dispersion of the $\tau$ estimates is
larger than for the other SED models and also larger than in the previous
examples. Thus, the overall accuracy is only slightly better than for the
1-MBB model, which however also corresponds to a lower resolution. When the
convolutions were performed using FFTs, the fits took of the order of one
minute for the 128$\times$128 pixel maps.

\subsection{Bonnor-Ebert sphere}  \label{results:BE}

We calculated synthetic surface brightness maps for a 2.6\,$M_{\sun}$
critically stable Bonnor-Ebert (BE) sphere \citep{Ebert1955,Bonnor1956}, with
an assumed gas temperature of 15\,K. However, in the subsequent radiative
transfer modelling the cloud is illuminated from the outside, with the
\citet{Mathis1983} interstellar radiation field (ISRF), and optionally from
the inside by a $10\,L_{\sun}$ point source that is modelled as a 10000\,K
black body. These result in continuous radial dust temperature gradients and
each LOS has emission from a different range of temperatures. In the outer
parts, the colour temperature approaches 18\,K. In the inner parts, the
temperature is below 14\,K for the externally heated and over 30\,K for the
internally heated case. The pixel size is set to 6$\arcsec$, with beam sizes
18$\arcsec$, 18$\arcsec$, 25$\arcsec$, and 36$\arcsec$ at 160, 250, 350, and
500\,$\mu$m, respectively. The subsequent SED fits assume a constant $\beta$,
which corresponds to the 160-500\,$\mu$m spectral index of the dust model
used in the radiative transfer calculations \citep{Compiegne2011}.

\begin{figure}
\centering
\includegraphics[width=9cm]{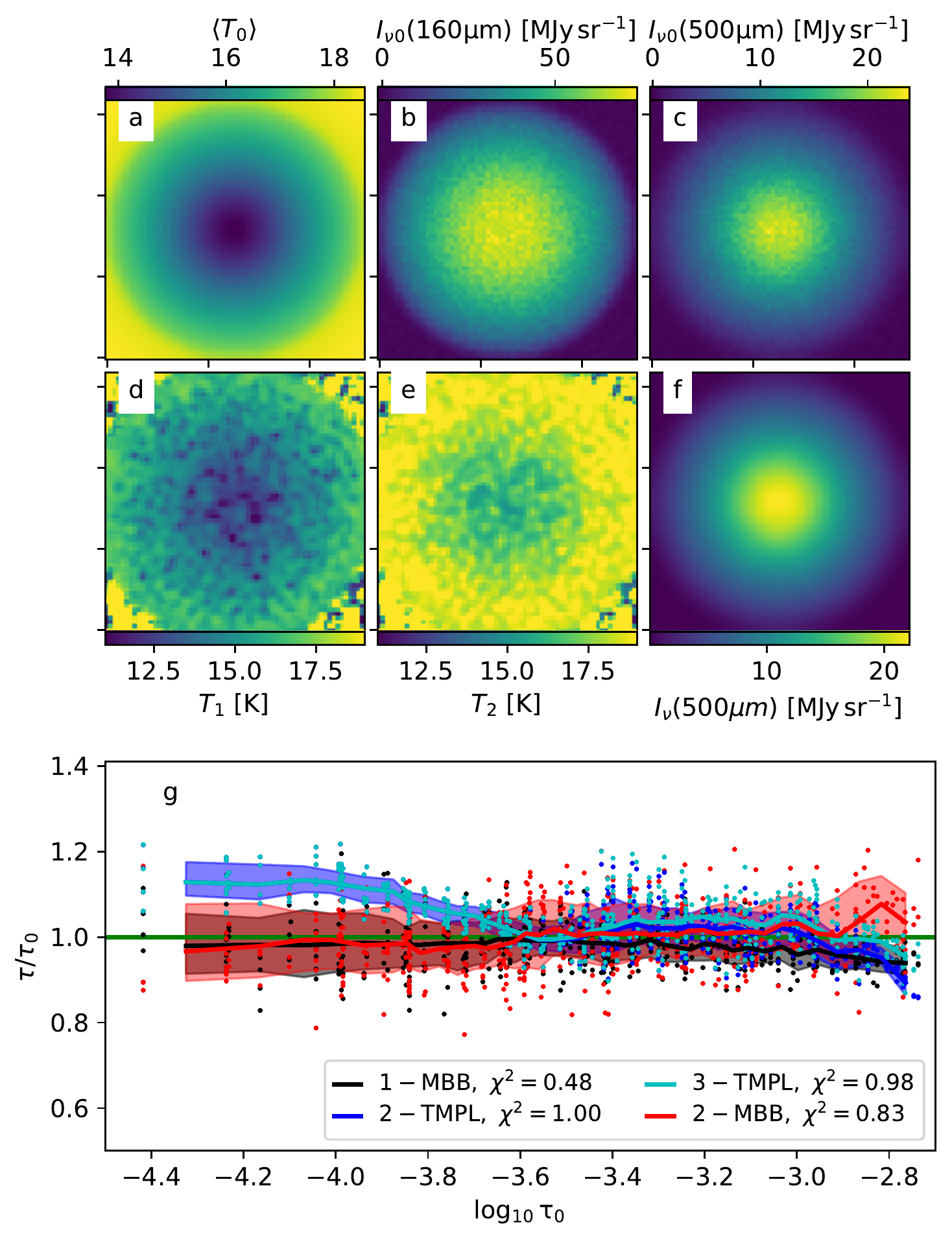}
\caption{
Fits on synthetic observations of an externally heated Bonnor-Ebert sphere.
Frames a-c show the mass-weighted average LOS temperature of the model cloud
and the calculated 160\,$\mu$m and 500\,$\mu$m surface brightness maps.
Frames d-e show the temperature maps and frame f shows the 500\,$\mu$m
predictions from the 2-MBB model. Frame g shows the ratio of the estimated
and true optical depth ($\tau/\tau_0$) as the function of $\tau_0$ for the
four cases listed in the legend. The solid lines are moving averages, the
shaded regions indicate the 1-$\sigma$ interval in the $\tau/\tau_0$ values,
and the individual pixels outside this interval are shown as dots.
}
\label{fig:TEST_BE_BG_BETA187}
\end{figure}

Figure~\ref{fig:TEST_BE_BG_BETA187} shows results for the externally heated
model. The 1-MBB model provides even surprisingly accurate results (at
36$\arcsec$ resolution). The 2-TMPL model uses $T_{\rm C}$ values of 14.5\,K
and 17\,K. These provide accurate estimates towards the cloud centre but
overestimate $\tau$ in the outer parts, where the dust temperatures are above
the $T_{\rm C}$ values. The 3-TMPL model has $T_{\rm C}$ values of 14, 15,
and 17\,K. Because the highest $T_{\rm C}$ is the same as in the 2-TMPL fit,
the results are similar in the outer parts of the cloud, but there is minor
improvement at the centre (corresponding to the added $T_{\rm C}$ value). The
width of the LOS temperature distribution appears to be sufficiently narrow
so that the 1-MBB model, by adapting to the LOS mean temperature, still
performs well compared to the 3-TMPL model. The 2-TMPL and 3-TMPL fits
enforced the positivity of the component weights but used no other priors.
Without the positivity requirement, the $\chi^2$ values would be lower but
especially the 2-TMPL result would include negative weights at some radial
distances.

The 2-MBB fits in Fig.~\ref{fig:TEST_BE_BG_BETA187} were also calculated
without priors. The estimates are mostly unbiased but show a relatively high
scatter. The corresponding noise in the temperature being visible in
Fig.~\ref{fig:TEST_BE_BG_BETA187}d-e, where temperatures are not well
constrained for individual pixels, at scales below the beam size. The low
$\chi^2$ value also indicates the possibility of some overfitting.

\begin{figure}
\centering
\includegraphics[width=9cm]{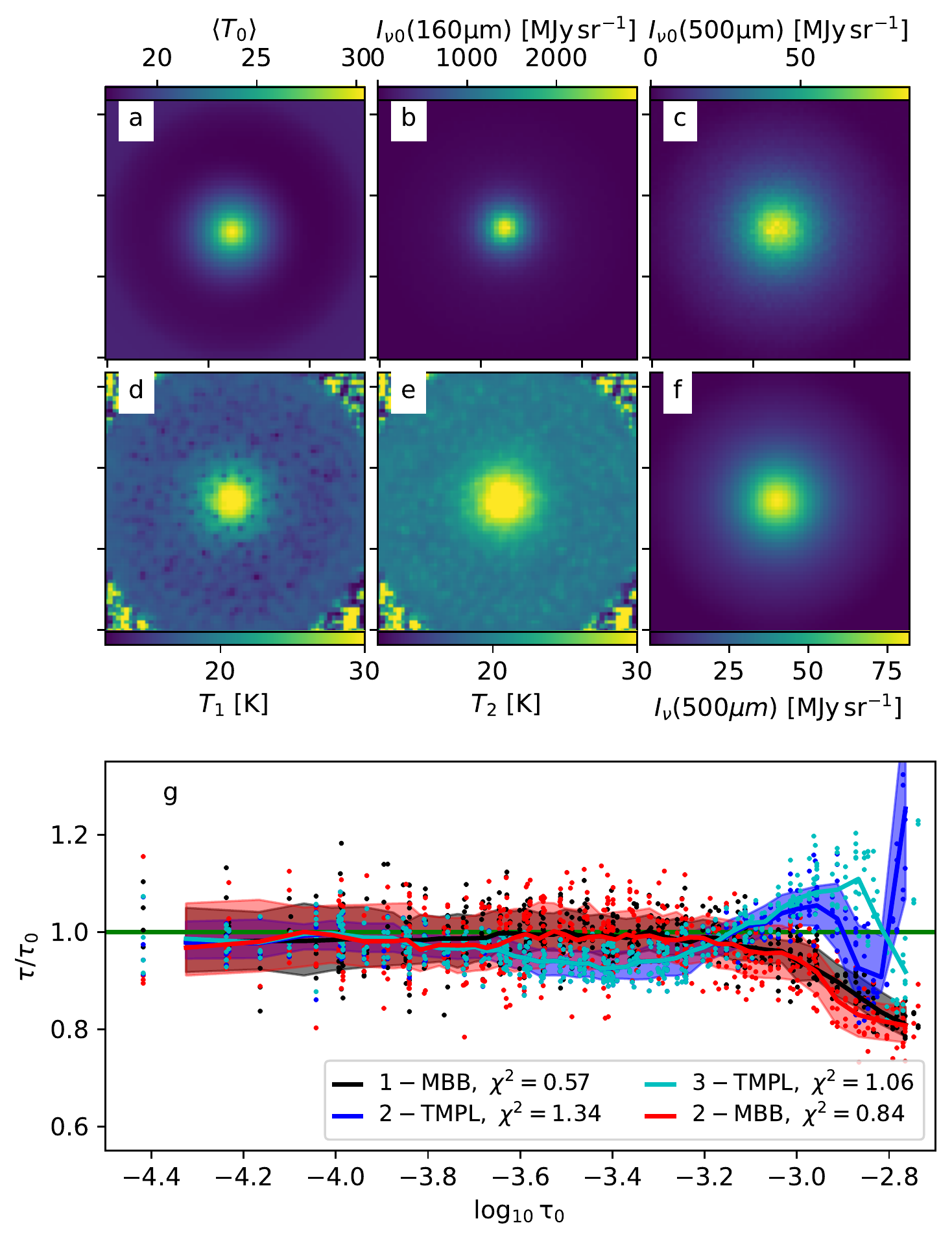}
\caption{
Same as Fig.~\ref{fig:TEST_BE_BG_BETA187}, but for a BE sphere with an additional
internal radiation source. The $T_{\rm C}$ values of the 2-TMPL and 3-TMPL
models are given in the text. The resolution is 36$\arcsec$ for the 1-MBB
model and 18$\arcsec$ for the other fits (for a nominal pixel size of
6$\arcsec$).
}
\label{fig:TEST_BE_PS_BETA187}
\end{figure}

Figure~\ref{fig:TEST_BE_PS_BETA187} shows the corresponding results for the
internally heated BE model. The errors of the 1-MBB fit increase to 20\%
towards the cloud centre. The selected $T_{\rm C}$ values are 18 and 25\,K
for the 2-TMPL model and 18, 23, and 29\,K for the 3-TMPL model. The 2-TMPL
model is unable to follow the rapid temperature changes in the centre, and
the oscillating errors result in some artefacts in the estimated radial
density profiles. The 3-TMPL model performs better at the very centre, but
the overall accuracy is not much better than for the 1-MBB model. 

The 2-MBB model shows increasing bias towards the cloud centre and the
$\tau/\tau_0$ curve actually follows that of the 1-MBB model predictions.
This is partly a coincidence, because the latter correspond to a factor of
two lower angular resolution. At the centre of the projected cloud, the 2-MBB
fit uses two temperatures that are both close to 30\,K and separated from
each other by some 4\,K. This is still quite small compared to the full range
of the actual LOS temperatures, which extend from $\sim 17$\,K to over 50\,K,
(the highest temperatures at scales below the beam sizes).

\subsection{Filament with stochastically heated grains}  \label{sect:filament}

We examined the optical depth estimates also as a function of the
wavelengths. This has already been studied extensively from the point of view
of observational noise and the LOS temperature variations
\citep{Shetty2009a,Shetty2009b,Malinen2011,Juvela2012_bananasplit,
Juvela2012_Tmix, Juvela2013colden, Juvela2013_TBmethods}. Below the main
questions are related not only to the relative performance of the $N$-MBB and
$N$-TMPL methods but also to the potential effect of stochastically heated
grains.

We used the filament simulations of \citet{Juvela2022b}.  The model is
discretised onto a Cartesian grid of 200$^3$ cells with a cell size of
0.0116\,pc. The model consists of a single filament with a Plummer-like column
density profile
\begin{equation}
N  = N_0 [ 1 + (r/R_0)^2 ]^{(1-p)/2}
\end{equation}
with $R_0$=0.0696\,pc and $p$=3. The column density perpendicular to the
filament major axis is set to $N({\rm H}_2)=10^{22}$\,cm$^{-2}$. The filament
is illuminated by a normal ISRF \citep{Mathis1983}. There can be an additional
15 700\,K point source with a 590\,$\rm L_{\sun}$ luminosity, which is is
placed 0.93\,pc in front of or to one side of the filament. The source is
0.93\,pc from the map centre, in a direction parallel to the filament main
axis. The dust properties, including those of stochastically heated grains,
are taken from \citet{Compiegne2011}.

We calculated 200$\times$200 pixel maps, where the pixel size corresponds to
the 3D discretisation and is set equal to 5$\arcsec$ (corresponding to a
distance of 479\,pc). Synthetic surface brightness maps were computed at 70,
100, 160, 250, 350, and 500\,$\mu$m with the assumed angular resolutions of
10$\arcsec$, 10$\arcsec$, 12$\arcsec$, 18$\arcsec$, 25$\arcsec$, and
36$\arcsec$, respectively. The wavelengths and resolutions are similar as in
\Herschel observations, except for the lower angular resolutions adopted for
the 70 and 100\,$\mu$m bands.

In the absence of the point source, the 1-MBB fits of the 160-500\,$\mu$m
observations give a narrow range of colour temperatures from 16\,K to
17.7\,K. Based on this, we selected temperatures $T_{\rm C}$=16\,K and 17\,K
for the 2-TMPL and $T_{\rm C}=$14, 16, and 18\,K for a 3-TMPL fit. When the
point source is added on the LOS towards the filament (0.93\,pc in front of
the filament and 0.93\,pc from the projected map centre), the range of 1-MBB
colour temperatures becomes 19.5-26\,K. If the point source is to one side of
the filament, the maximum exceeds 40\,K but only towards the source position,
in a region of relatively low column density. In these cases, we used $T_{\rm
C}$=21 and 25\,K for the 2-TMPL fits and $T_{\rm C}$=20, 23, and 26\,K for
the 3-TMPL fits. All surface-brightness observations contain 3\% relative
noise.

\begin{figure}
\centering
\includegraphics[width=9cm]{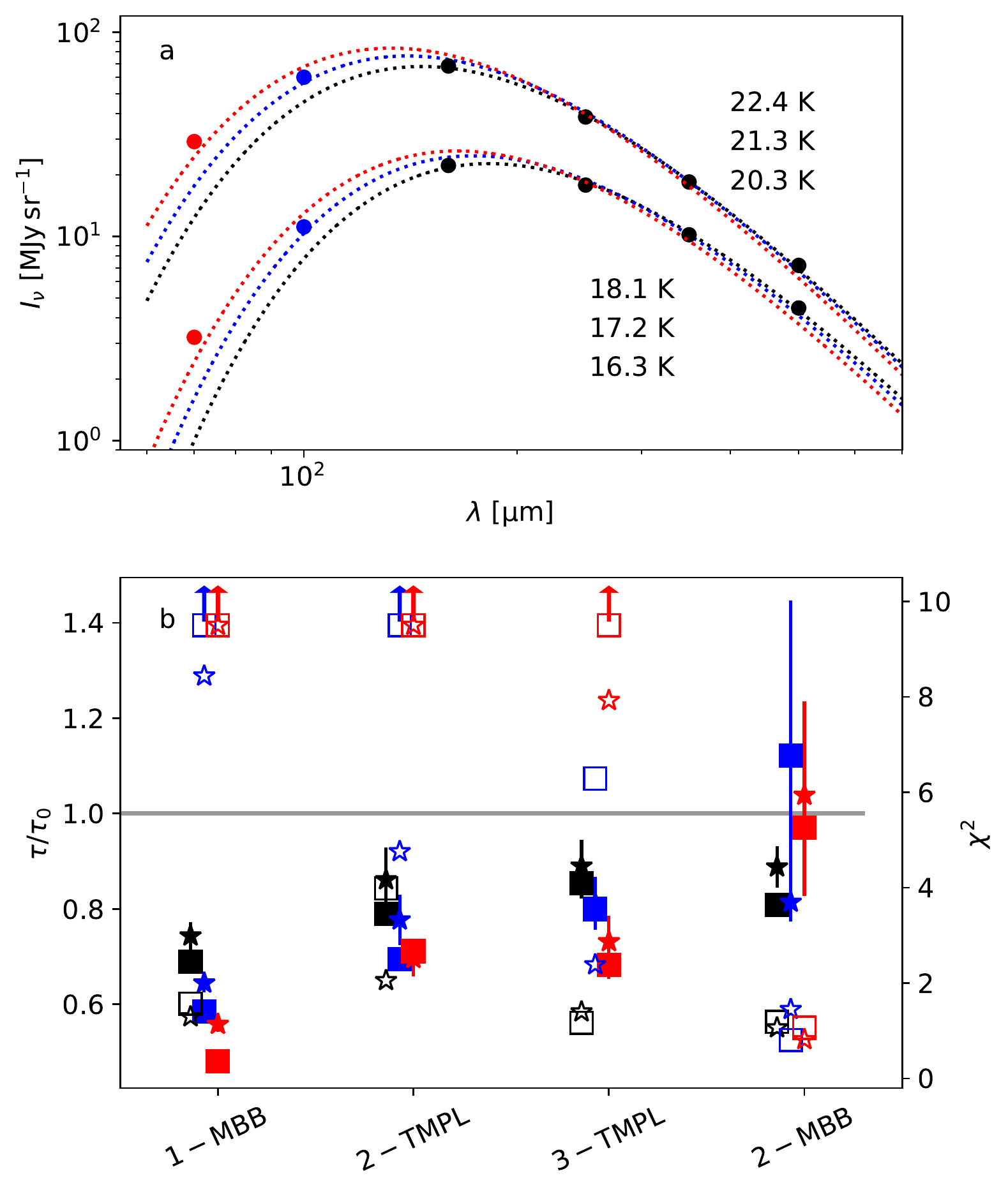}
\caption{
SED fits of the filament models. Frame a shows the SEDs towards the map
centre and the 1-MBB fits to the 4-6 longest wavelength bands. The lower set
of symbols and curves corresponds to a model without a point source, and the
upper ones to a model with a point source on the LOS towards the filament but
at 0.93\,pc in front of the filament. The colour temperatures of the 1-MBB
fits are quoted in the frame.
Frame b shows the ratio of recovered and true peak optical depths (left axis,
filled symbols) and the average $\chi^2$ per data point over the whole
map (right axis, open symbols). The x-axis indicates the four models. The
square and star symbols are, respectively, for the externally heated model and
the model with a point source. The black, blue, and red symbols correspond to
fits that extend to shortest wavelengths of 160, 100, and 70\,$\mu$m,
respectively. The error bars show the standard deviation of the optical depth
estimates along the filament spine.
}
\label{fig:TEST_FILAMENT_STAT}
\end{figure}

Figure~\ref{fig:TEST_FILAMENT_STAT}a shows SEDs towards the map centre. These
correspond to a single pixel and models without the point source or with the
point source on the LOS towards the filament centre (but offset from the 3D
centre of the filament). The fits show the systematic increase of the colour
temperatures as the 1-MBB fits are extended to higher frequencies. The lower
frame shows how the SED fits translate to $\tau$ estimates. The optical
depths tend to be underestimated, and this also applies to the 2-TMPL and
3-TMPL models. The systematic error is largest for the 1-MBB model, roughly
equal for the 2-TMPL and 3-TMPL models, and smallest for the 3-TMPL model.
Like the average colour temperature, the bias of the $\tau$ estimates tends
to increase as shorter wavelengths are included in the fit. The exception are
the 2-MBB fits, where the statistical errors dominate. The error bars in
Fig.~\ref{fig:TEST_FILAMENT_STAT}b correspond to the $\tau$ estimates over a
0.7\,pc long section of the filament spine. In models containing a point
source, this section also includes the projected location of the point
source.

Figure~\ref{fig:TEST_FILAMENT_STAT}b shows that the $\chi^2$ values per
data point of the 2-MBB fits are of the order of one for all band
combinations. In the 1-MBB fits, the increase in the number of bands is
associated with a rapid increase of $\chi^2$ values (as indicated by the SEDs
in frame a). A similar trend exists for the 2-TMPL model and to a lesser
extent for the 3-TMPL model, these of course depending on the chosen $T_{\rm
C}$ values.

\subsection{MHD cloud simulation}  \label{results:IRDC}

As the final test we examined a MHD cloud simulation. The full model volume is
(4\,pc)$^3$ in size, but we used only the subvolume of (1.84\,pc)$^3$ that
contains the densest structures. The column densities reach values above
$N({\rm H}_2)\sim 10^{23}$\,cm$^{-2}$, while the average volume density is
still below 10$^3$\,cm$^{-3}$. The simulation includes a few hundred embedded
sources that increase the temperature variations. Details of the MHD
simulation can be found in \citet{Haugbolle2018} and of the radiative transfer
modelling in \citet{Juvela2022a}. Because the model includes a filamentary
structure reminiscent of a small infrared dark cloud (IRDC), we refer to the
model as the IRDC cloud. 

The synthetic observations consist of maps at 160\, 250\, 350, and
500\,$\mu$m. The effect of stochastically heated grains is not included. For
the SED fitting, the main challenges are the large map sizes and the
temperature variations associated with the cores. The 3D dust temperatures
vary from less than 10\,K in the cold cores to up to $\sim$100\,K in the hot
cores, but the range of the observed colour temperatures is naturally smaller.
The surface brightness maps have 1875$\times $1875 pixels, and the assumed
beam sizes are 2.3 , 2.3, 3.1, and 4.5 pixels at 160\, 250\, 350, and
500\,$\mu$m, respectively. We use the same wavelengths as in previous tests,
although the above scaling would correspond to a source distance of only
25\,pc for a telescope of the \Herschel size. The observational noise is 3\%
of the observations.

Below we compare the 1-MBB results to 3-TMPL and 3-MBB fits and then,
with additional priors, to $N$-TMPL and $N$-MBB fits with $N>3$. We also
look at the $N$-TMPL fits made in Fourier space separately, because those adopt a
different noise model and are more limited in the use of priors. We are also
interested in the relative run times.

\subsubsection{IRDC cloud and 2-MBB and 4-TMPL fits} \label{sect:IRDC4}

\begin{figure}
\centering
\includegraphics[width=9cm]{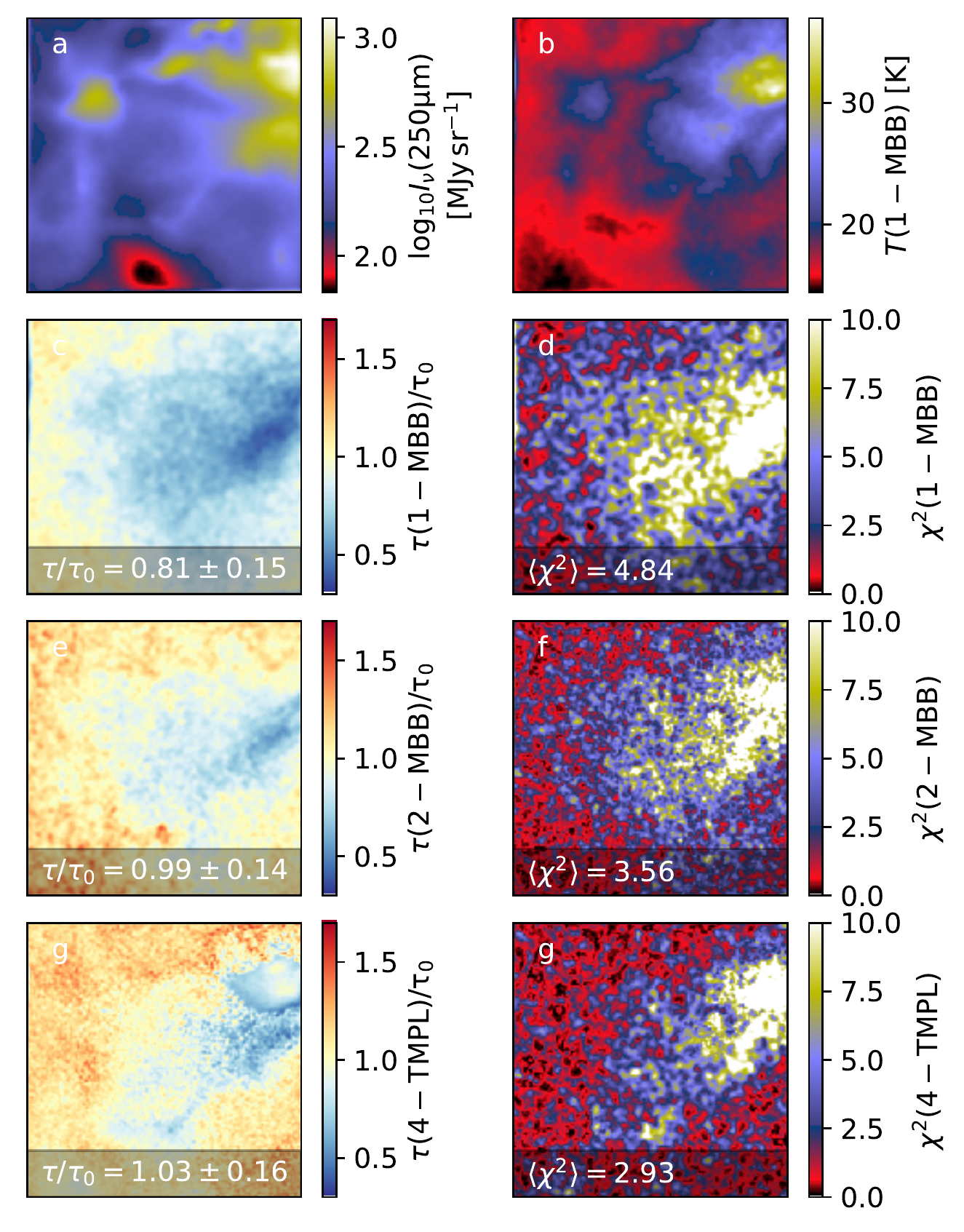}
\caption{
IRDC observations fitted with 1-MBB, 2-MBB, and 4-TMPL models. Frames a-b show
the 250\,$\mu$m intensities and colour temperatures from the 1-MBB fit. The
second row shows the 1-MBB $\tau$ estimates relative to the correct values
(frame c) and the $\chi^2$ values (frame d). The third and fourth rows show
the corresponding data for the 2-MBB and 4-TMPL fits. The plots cover a
256$\times$256 pixel area of the IRDC cloud model.
}
\label{fig:TEST_IRDC_N3}
\end{figure}

Figure~\ref{fig:TEST_IRDC_N3} compares the 1-MBB fit to the results from 2-MBB
and 4-TMPL fits for a 256$\times$256 pixel area of the full IRDC maps. Even
this small region includes more than two orders of magnitude differences in
surface brightness, with colour temperatures ranging from $\sim$14\,K to over
30\,K. The analysis uses a $\beta$ value that is correct for the
160-500\,$\mu$m interval. The 4-TMPL fit uses $T_{\rm C}$ values of 15, 18,
22, and 27\,K. The models were fitted without direct priors, except for the
temperatures of the 2-MBB fit being limited to values above 8\,K.

In the 1-MBB fits, the average ratio between the estimated and true optical
depths is $\langle \tau/\tau_0\rangle=0.81$ and the standard deviation
$\sigma(\tau/\tau_0)=15$\%. The optical depths are always underestimated, and
the errors reach $\sim$60\% in a region that contains the superposition of
warm extended emission and a more compact and colder filamentary structure.
The $\chi^2$ values are mostly well above one and in this case also correlated
with the magnitude of the systematic $\tau$ errors.

For the 2-MBB and 4-TMPL fits, the morphology of the $\tau$ and $\chi^2$ maps 
is similar to the 1-MBB case. The bias of the $tau$ estimates is somewhat
lower, and also the $\chi^2$ values are lower, especially considering that
they refer to up to a factor of two higher angular resolution. The maximum
errors are still almost $50$\% but they are limited to smaller regions. The
$T_{\rm C}$ values of the 4-TMPL model were chosen so that they result in
accurate $\tau$ estimates. However, the results are not very sensitive to the
precise $T_{\rm C}$ values, as long as they cover the range of colour
temperatures in the 1-MBB fit.

\subsubsection{IRDC cloud and 11-TMPL and 4-MBB fits} \label{sect:IRDC11}

When $N$ is increased further, the $N$-TMPL and $N$-MBB fits require
additional constraints. When $\beta$ is kept constant, a 4-MBB model has eight
free parameters and an 11-TMPL model has 11 free parameters, both thus much
larger than the number of the observed wavelengths. 

\begin{figure}
\centering
\includegraphics[width=9cm]{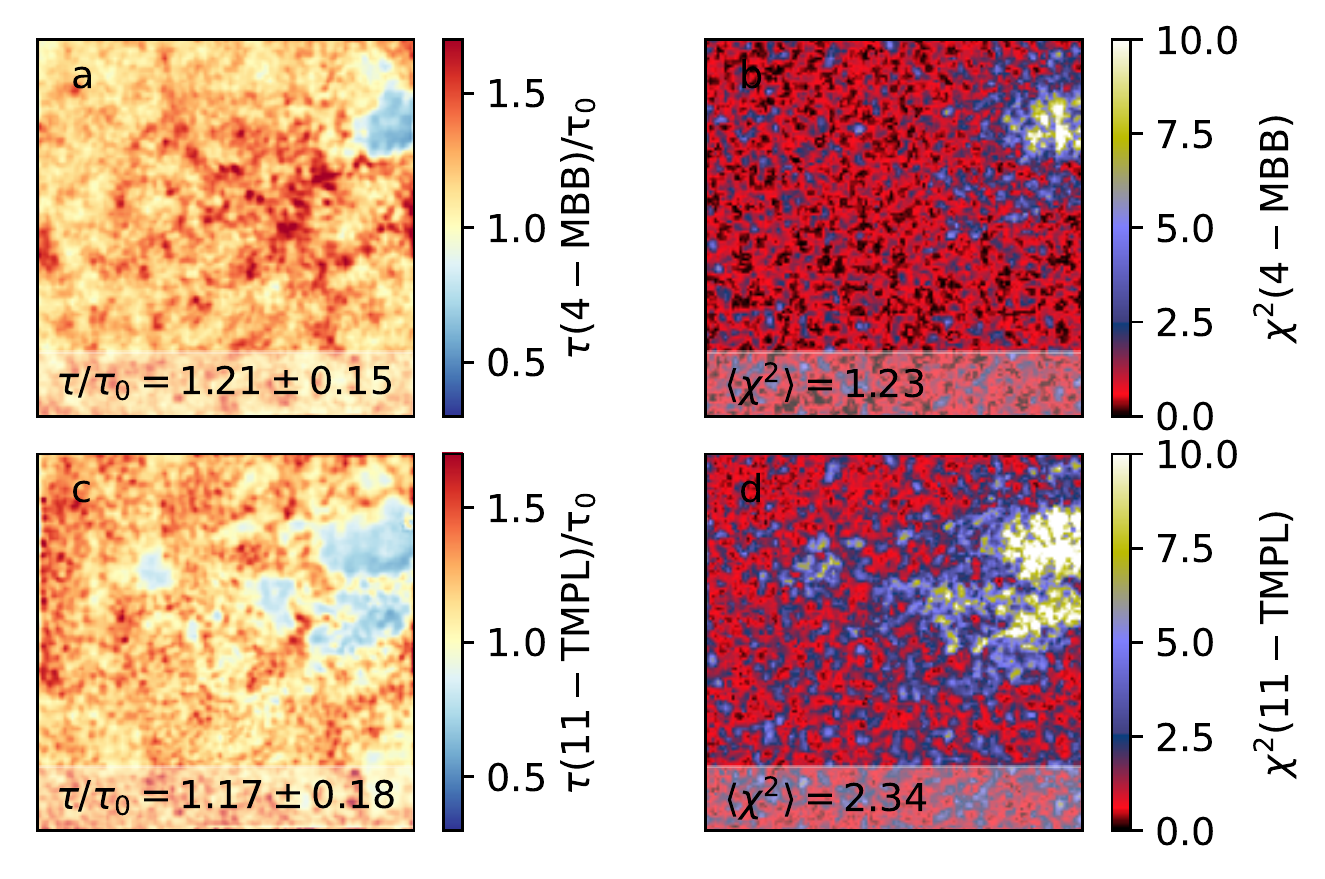}
\caption{
Fits of IRDC model maps with 4-MBB and 11-TMPL models. The corresponding data
for the 1-MBB fit are shown in Fig.~\ref{fig:TEST_IRDC_N3}.
}
\label{fig:TEST_IRDC_N11}
\end{figure}

Figure~\ref{fig:TEST_IRDC_N11} shows results for the same area as in
Fig.~\ref{fig:TEST_IRDC_N3}. The $T_{\rm C}$ values of the 11-TMPL model are
placed logarithmically between 12\,K to 40\,K. The temperature prior is $T\sim
N(25\,{\rm K}, 3\,{\rm K})$ for both the 4-MBB and 11-TMPL fits. The large
$\chi^2$ values in the upper right hand part show that the fits do not yet
correctly describe the complex temperature distributions of this area.
Elsewhere in the maps the $\chi^2$ values are at or below $\sim 1$, but low
$\chi^2$ values do not translate to more accurate $\tau$ estimates. The
$\tau/\tau_0$ ratios are now more constant but there is also higher positive
bias. The use of different apparently reasonable priors result in $\langle
\tau/\tau_0\rangle$ varying by more than 10\%. In particular, weaker priors
tend to lead to larger temperature fluctuations and larger positive bias in
$\tau$. In all cases, the dispersion in the $\tau/\tau_0$ ratios is equal or
larger than in the 1-MBB case (not accounting for the difference in the
angular resolutions).

The same fits for the whole 1875$\times$1875 pixel maps are shown in
Appendix~\ref{app:1875}. The average statistics are similar to the previous
image, although the lower average column density results in lower average
bias, especially in the case of the 1-MBB fit. However, there are also regions
of stronger internal heating, where all the fitted SED models can
underestimate the optical depths by up to 60\%.

\subsubsection{IRDC cloud and $N$-TMPL fits in Fourier space} \label{sect:IRDC_FOU}

Fits of the $N$-TMPL model in Fourier space (i.e. $N$-TMPL-F fits) are
attractive because of the faster computation, even if it only allows limited
use of priors. The error estimates are also required to be constant over the
map, and we set them equal to the mean of the 3\% relative uncertainties. We
also add to the minimised $\chi^2$ function a penalty
\begin{equation}
P = \xi \sum_i W_i^2.
\label{eq:pen2}
\end{equation}
This biases the solution against individual high $W_i$ and, more importantly,
makes it unlikely that an observation would be fitted as a sum of positive and
negative terms instead of a smaller number of smaller positive components.

\begin{figure*}
\centering
\includegraphics[width=18cm]{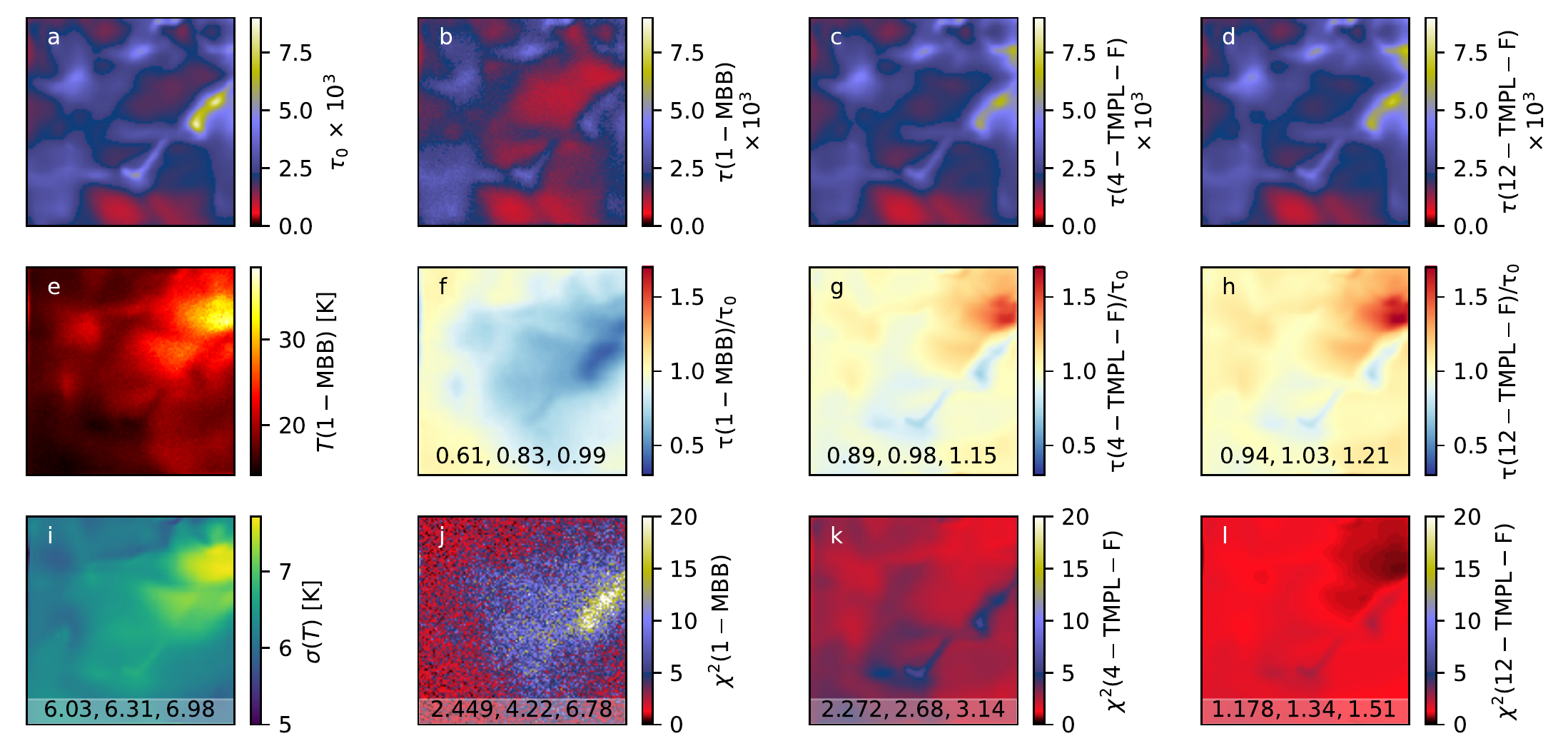}
\caption{
IRDC observations fitted in Fourier space. Frame a shows the true optical
depth and frame e the colour temperature from the 1-MBB fit. Frames b-d show
the optical depths estimated with the 1-MBB, 4-TMPL, and 12-TMPL fits,
respectively. The corresponding ratios of estimated and true optical depths
are shown in frames f-h and the $\chi^2$ values in frames j-l. Frame i shows
the standard deviation of the temperature values from the 12-TMPL fit. The
values quoted in the frames are the 10\%, 50\%, and 90\% percentiles of the
plotted quantity. The maps cover the same 256$\times$256 pixel area as in
Fig.~\ref{fig:TEST_IRDC_N3}.
}
\label{fig:TEST_IRDC_3_FOU_256}
\end{figure*}

Figure~\ref{fig:TEST_IRDC_3_FOU_256} compares the 1-MBB results to the
4-TMPL-F and 12-TMPL-F fits. The $T_{\rm C}$ values were placed
logarithmically in the range 11-45\,K. The fits were done without direct
priors on the temperature. However, the 12-TMPL-F fit included a small penalty
according to Eq.~(\ref{eq:pen2}), with a $\xi$ values such that the
effect on the final $\chi^2$ was $\la 1$\%. 

The 1-MBB results are identical to the previous tests, apart from the
difference in the adopted error estimates, and still underestimate the true
optical depth by up to 60\%. Although the differences are not very large, the
4-TMPL and 12-TMPL fits are here more accurate, with a more symmetric error
distribution around the correct value. The average $\langle \tau/\tau_0
\rangle$ value of the 12-TMPL-F fit varies by $\pm 0.1$ units if the number of
components is changed between 10 and 20, the lower limit of the $T_{\rm C}$
values between 10\,K and 14\,K, and the upper limit between 30\,K and 40\,K.
The 4-TMPL-F fit is also relatively robust with respect to selected $T_{\rm
C}$ grid. The 10\% and 90\% percentiles of the $\tau/\tau_0$ ratios (shown in
Fig.~\ref{fig:TEST_IRDC_3_FOU_256}) are very similar for the 4-TMPL and
12-TMPL fits, in spite of the factor of two difference in the $\chi^2$ values.

The upper right corner includes a region where the 4-TMPL-F and 12-TMPL-F fits
overestimate $\tau$ by almost $\sim$60\%. If the $T_{\rm C}$ grid is extended
to 55\,K, the maximum errors of the 12-TMPL-F fit are confined to a more
symmetric range of [-35\%, +31\%]. This is in agreement with the maximum LOS
temperatures being higher than the maximum 1-MBB colour temperatures of
$\sim$35\,K. When the highest $T_{\rm C}$ is 35\,K, some components around
$T_{\rm C}\sim 30$\,K have negative weights in the area of the largest $\tau$
errors, in spite of the use of Eq.~(\ref{eq:pen2}) penalty. The solution is
thus partly unphysical, although, to the extent that the emission is in the
Rayleigh-Jeans regime, this does not translate to large errors in the $\tau$
estimates. When the $T_{\rm C}$ grid is extended to higher temperatures,
practically all weights remain positive.

The fits performed with the same parameters for the whole 1875$\times$1875
pixel maps are shown in Appendix~\ref{app:1875}. Due to the lower average
column density, the mean bias of the 1-MBB estimates is there closer to zero
(Fig.~\ref{fig:1875fou}), and also the 10-90\% intervals of the $\tau/tau_0$
ratios are similar between the fits. There is a number of hot sources where
$\tau$ is underestimated by all three methods.

\section{Discussion} \label{sect:discussion}

Most studies model the FIR dust emission with a single MBB function (1-MBB),
and the observations (the noise level and number of observed wavelengths) also
may not allow much more complex analysis. The main problem of the
single-temperature model is the underestimation of column densities
\citep{Shetty2009a, Juvela2012_Tmix}. The bias depends on the width of the
temperature distribution, which makes SF regions with hot and cold cores
particularly challenging. We have examined in this paper the use of
multi-component SED models, using observations at their native resolution. We
discuss the main findings below.

\subsection{Basic tests}

When the number of free parameters is less than the number of observed
wavelengths, the SED fits can be done without additional constraints. The use
of prior distributions may still be beneficial, especially when the data have
low S/N.

In the tests of Sect.~\ref{results:toy} and Sect.~\ref{results:BE}, the 2-MBB
fits provided only small improvement over the 1-MBB fit. They resulted in
lower bias in the $\tau$ estimates but with a larger statistical noise. The
$\chi^2$ values were not directly correlated with the precision of the $\tau$
estimates. The low number of SED components was more of a problem for the
2-TMPL model, because the $T_{\rm C}$ values remain fixed and, unlike in the
$N$-MBB models, not adjusted for each LOS separately. The errors depended on
the distance between the $T_{\rm C}$ values and the actual source temperatures
and increased rapidly outside the $T_{\rm C}$ range. This is a problem, since
the full range of temperatures may not be known a priori and is difficult to
estimate from observations. On the other hand, the errors may have been
exaggerated in Sect.~\ref{results:toy}, because of the discrete temperature
values of the synthetic observations. In real sources the temperature
distribution is always wider, leading to less sharp changes in the bias.

In Sect~\ref{results:BE}, the internally heated Bonnor-Ebert sphere could not
be fitted properly with just two temperature components. The absolute errors
of the 3-TMPL were below $\sim$20\% but showed rapid oscillations as the mean
LOS temperature moved across the $T_{\rm C}$ grid. When the true solution is
not available, such oscillation cannot be recognised, and this could have a
significant effect on the derived density profiles. When beam convolutions are
included in the SED models, the 2-MBB model also showed some problems with
temperature oscillations at scales below the beam size. When the predictions
of such a model are convolved to the observed resolution, the SED fits may be
perfect but the predicted column densities will show positive bias that
depends especially on pixels with the coldest temperatures. This might require
additional constraints on the smoothness of the solution, provided that
temperature field in the source is indeed assumed to be smooth at the observed
resolution. Such priors would also affect the effective angular resolution of
the obtained $\tau$ maps.

\subsection{Filament with stochastically heated grains}

In Sect.~\ref{sect:filament}, we examined the dependence of $\tau$ estimates 
on the observed wavelengths in the 70-500\,$\mu$m range. It is well known that
the inclusion of shorter wavelengths tends to increase the estimated
temperatures \citep[e.g.][]{Shetty2009a}. This is based on the warmer dust
having a higher contribution (relative to its mass) to the observed
intensities. We had additional effects from stochastically heated grains that
contributed significantly only to the shortest wavelengths. In a strong
radiation field, the 70\,$\mu$m emission could be emitted by warm large
grains, but in our model the stochastically heated grains dominated the
70\,$\mu$m observations and had a clear contribution to the 100\,$\mu$m
emission.

When the fitted wavelength range was extended from 160-500\,$\mu$m to
70-500\,$\mu$m, the optical depths were increasingly underestimated. The error
was always more than 20\% in the 1-MBB fits and mostly around 20\% for the
other fits. The 2-MBB fit was an exception, showing no clear bias (partly
because of the larger statistical errors). There was no large difference
between filaments that were illuminated by an isotropic radiation field or also by
a nearby star. The main effect is that a higher radiation field results in
higher temperatures, which always tends to decrease the systematic errors by
moving the observations further into the Rayleigh-Jeans regime.

\subsection{The IRDC model}

The IRDC model represented a complex star-forming cloud with a large range of
temperatures and column densities. Therefore, the 1-MBB fits showed
significant errors with the optical depths being underestimated by up to 60\%.
The 2-MBB and 4-TMPL models resulted in some improvement, although this
required a careful selection of the $T_{\rm C}$ values for the 4-TMPL model.
However, as long as the number of free parameters is not larger than the
number of observed wavelengths, the solution is more or less well constrained,
and there is some benefit from using SED model that describe not only the
average temperature but also the temperature variations within the beam.

We also tested SED models with a larger number of free parameters. These did not
produce clear benefits (apart from lower $\chi^2$ values). On the contrary,
these require stronger priors that risk biasing the results. The noise in
multi-component models tends to increase the effective temperature dispersion
of the SED model, which always biases the $\tau$ estimates upwards. The run
times scales for both $N$-TMPL and $N$-MBB models approximately linearly with
the number of SED components. The $N$-TMPL fits can be done faster in Fourier
space but, as discussed in Sect.~\ref{sect:methods}, the method has its
limitations. The error estimates are (at least in our implementation) limited
to a single value per, instead of individual error estimates for each pixel
and frequency. One also cannot use direct priors for the weights $W_i$,
although it is possible to give priors for the Fourier amplitudes, as was done
in Sect.~\ref{sect:IRDC_FOU} to bias against negative weights. In
Fig.~\ref{fig:TEST_IRDC_3_FOU_256}, both the 4-TMPL and 12-TMPL fits resulted
in clear improvement over the 1-MBB results. However, when the same parameters
were used to analyse the full 1875$\times$1875 pixel map, the improvements
were less clear. The differences between the different SED models can in
itself be a useful diagnostic, giving indications of the model errors that are
not covered by the normal fit error estimates.

\subsection{Degeneracies in multi-component SED fits} \label{sect:degeneracies}

One factor affecting the systematic and statistical noise of the
multi-component fits is the beam convolution. The multi-component fits $N$-MBB
and $N$-TMPL try to estimate $T$ and $\tau$ at the highest observed
resolution. Since the resolution varies with wavelength, the best intensity
information is available on a higher resolution than the information on the
SED shape. There are even weaker constraints for the SED model parameters in
individual map pixel, because only the averages over the beam are directly
constrained. This was most directly visible in the $N$-MBB fits where, as the
small-scale noise.

\begin{figure}
\centering
\includegraphics[width=9cm]{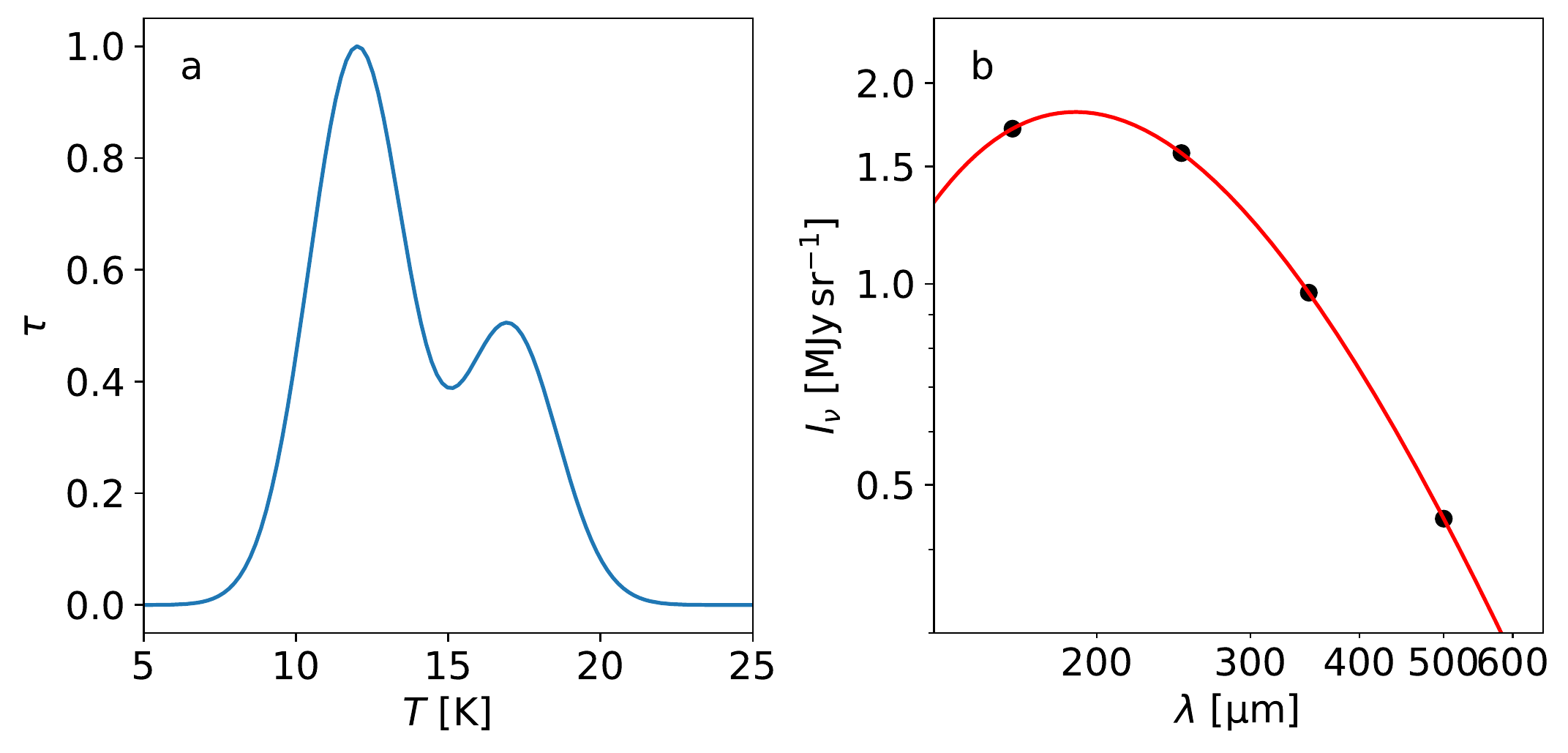}
\caption{
Simulated single-pixel observation. Frame a shows the optical depth
distribution as a function of temperature, and frame b shows the SED, with
measurements at 160, 250, 350, and 500\,$\mu$m.
}
\label{fig:toy_model_3a}
\end{figure}

\begin{figure}
\centering
\includegraphics[width=9cm]{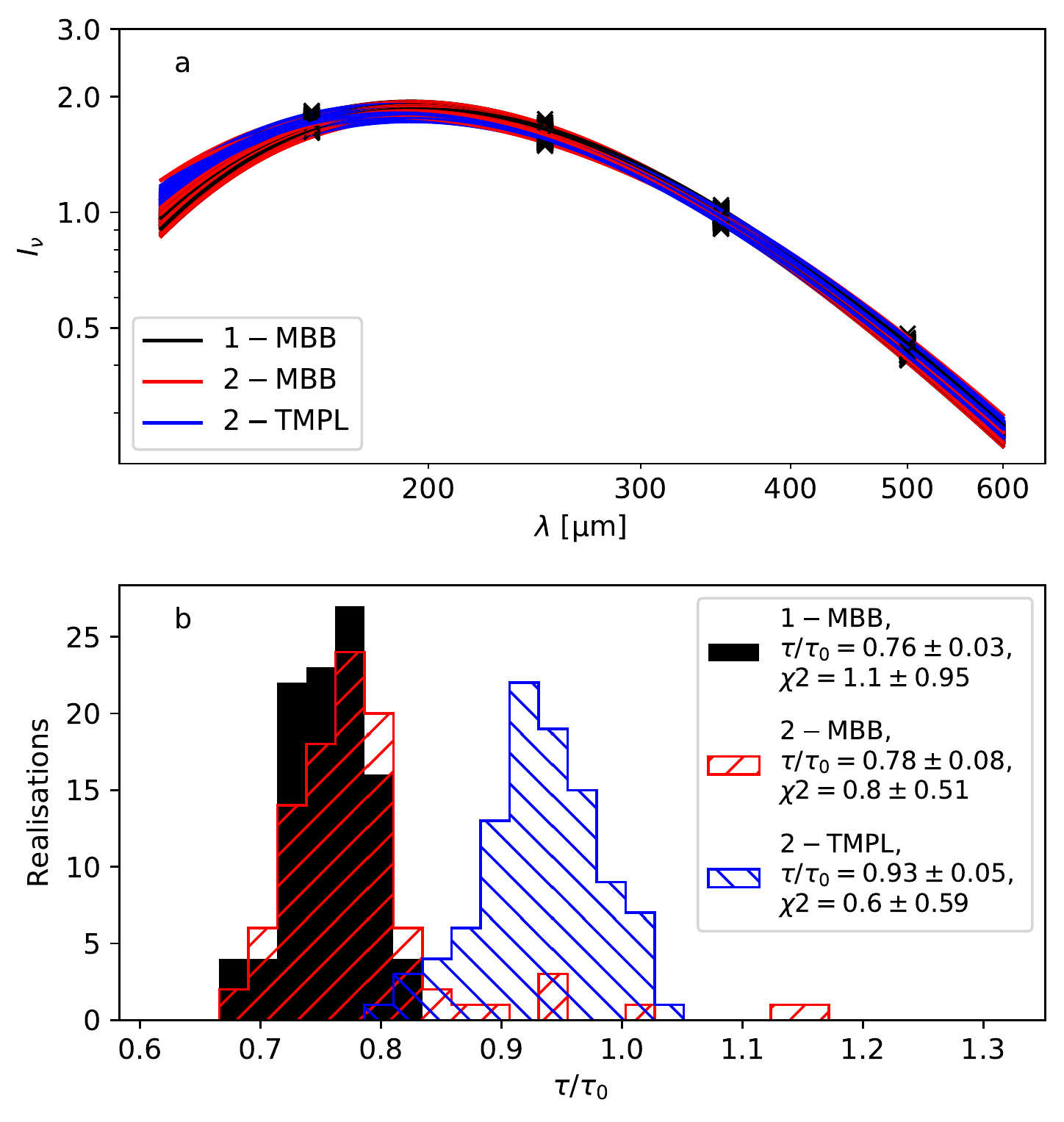}
\caption{
Results from the fitting different noise realisations of the
Fig.~\ref{fig:toy_model_3a} observations. Frame a shows a few examples of the
observations (black crosses) and individual fits (solid curves). Frame b-d
show the distributions of $\tau$ estimates (100 noise realisations) for the
1-MBB, 2-MBB, and 3-MBB models, respectively. Each frame quotes the mean value
and the standard deviation for the estimates relative to the true value,
$\tau/\tau_0$.
}
\label{fig:toy_model_3b}
\end{figure}

Even without the added complication of beam convolution, multi-component fits
suffer from degeneracies between the fitted parameters. This can be
demonstrated easily even for observations of a single pixel.
Figure~\ref{fig:toy_model_3a} shows the SED for a LOS dust temperature
distribution that consists of Gaussian components at 12\,K and 17\,K.
Figure~\ref{fig:toy_model_3b} shows the corresponding distributions of the
estimated optical depths for 100 noise realisations. The average bias of the
resulting $\tau$ estimates is 25\% for the 1-MBB model, 20\% for the 2-MBB
model, and 10\% for the 2-TMPL models. If 2-MBB fits are performed without
priors, a significant fraction of fits include one component with very low
temperature (fitting some noise feature), which then leads to $\tau$ being
overestimated. The error can reach orders of magnitude, if temperatures
$T\ll10$\,K are allowed. In Fig.~\ref{fig:toy_model_3a}, we use a temperature
prior $T\sim N \rm (15\,K, 3\,K)$, which is still not enough to prevent
occasional 50\% overestimation of $\tau$. On the other hand, if we were
observing a field that potentially includes cold cores, the above prior would
bias their $\tau$ estimates in the other direction, leading the optical depth
of cold cores to be underestimated.

The 2-TMPL model performs well in Fig.~\ref{fig:toy_model_3a}, because the
$T_{\rm C}$ values were chosen to match the peaks of the true temperature
distribution. The results change only little if the $T_{\rm C}$ values are
both moved 1\,K higher or lower. However, the ratio $\langle \tau/\tau_0
\rangle$ drops to 0.80 if the $T_{\rm C}$ values are both moved 1\,K further
apart, and increases to $\langle \tau/\tau_0 \rangle$=1.16 if the $T_{\rm C}$
values are both moved 1\,K closer to each other (following the trends seen in
Sect.~\ref{results:toy}). The most accurate solution also corresponds to the
lowest average $\chi^2 $ value, $\langle \chi^2 \rangle$=0.5 (compared to
$\langle \chi^2 \rangle \sim$1 for the wider and the narrower $T_{\rm C}$
separations). For all the $T_{\rm C}$ combinations mentioned above, the
$\chi^2$ distribution (over different noise realisations) extends from $\chi^2
\sim 0$ to $\chi^2\ \sim 2$ and above. Therefore, for an individual
measurement, the best model cannot be reliably selected based on the $\chi^2$
value alone. While 2-TMPL and 2-MBB may have smaller bias, they are not a
priori much more accurate in predicting correct $\tau$ values. For a
collection of observations, the data may contain enough information for the
selection of the most appropriate priors (manually or by using hierarchical
statistical models). The observations should thus conform to a common,
preferably narrow range of SED parameters. Otherwise the effect of priors will
be diluted or, alternatively, they may bias the results for sources outside
the main parameter distribution.

\begin{figure}
\centering
\includegraphics[width=9cm]{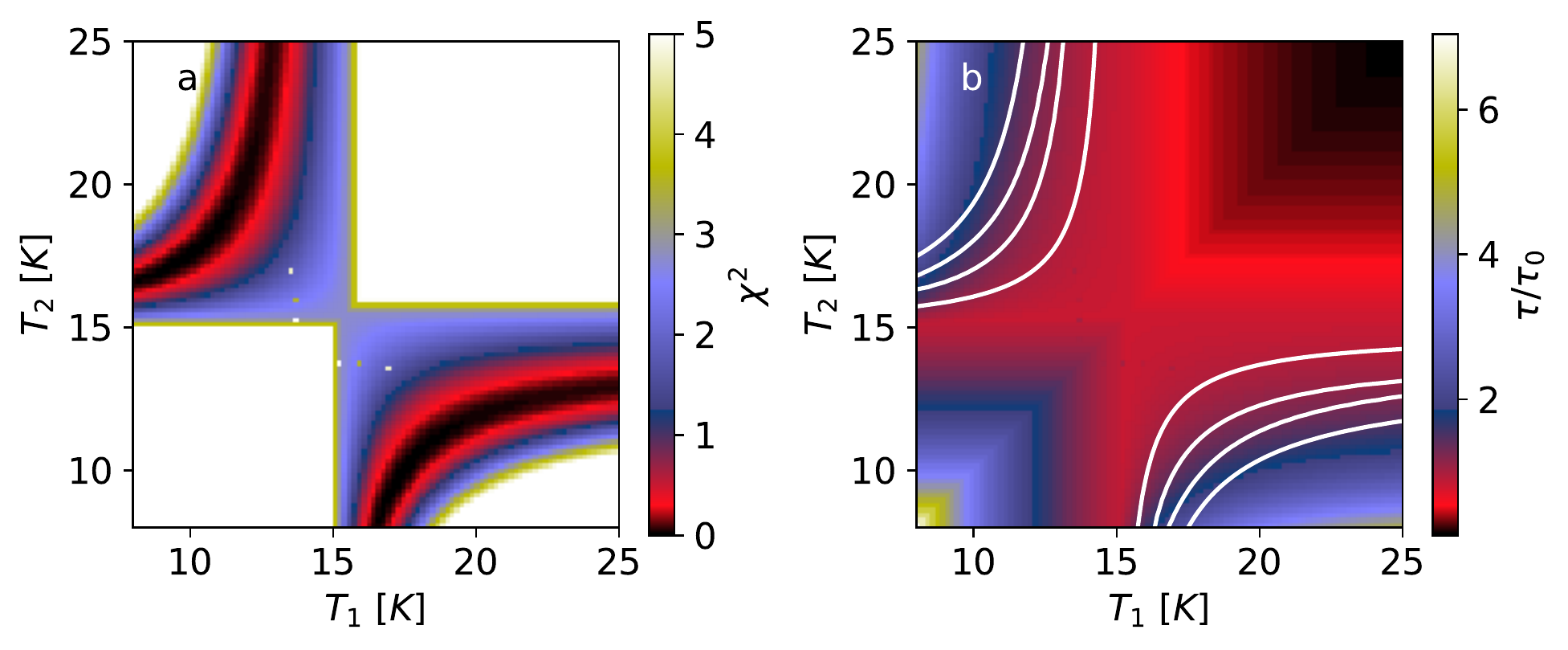}
\caption{
Values of $\chi^2$ (frame a) and $\tau/\tau_0$ (frame b) as functions of the
assumed temperatures $T_{\rm 1}$ and $T_{\rm 2}$ of the fitted SED templates.
The observations correspond to one noise realisation selected from the
simulations in Fig.~\ref{fig:toy_model_3b}. Frame b includes contours at
$\chi^2=0.1$ and $\chi^2=1$.
}
\label{fig:exam_bad_fits}
\end{figure}

When the number of observed wavelengths is small, both the 2-TMPL and 2-MBB
fits can result in $\chi^2$ values below one. This suggests that the $\tau$
estimates may be significantly affected by the noise realisation (as well as 
any potential priors). The results of Sect.~\ref{results:toy} showed that a
low $\chi^2$ value even of the 1-MBB model is not a guarantee of accurate
column density estimates. When the number of free parameters is increased,
values $\chi^2<1$ can be obtained with different parameter combinations. This
is illustrated by Fig.~\ref{fig:exam_bad_fits} that is based on one noise
realisation from Fig.~\ref{fig:toy_model_3b}. This is worse than an average
case but illustrates the potential for parameter degeneracies.
Figure~\ref{fig:exam_bad_fits}a shows the $\chi2$ values as a function of the
temperatures of two components. The intensities of the components are
optimised separately for each pixel in the figure (i.e. for every temperature
combination). The plot is reminiscent of the degeneracy of $T-\beta$ fits,
with long curved valleys where the minimum $\chi^2$ values remain almost
constant \citep[cf.][]{Chastenet2021}. Figure~\ref{fig:exam_bad_fits}b shows
that fits with $\chi^2 = 1$ correspond to a range of solution, where the
difference between the highest and lowest $\tau$ estimates is more than a
factor of two. In 2-MBB fits, any of these temperature combinations could be
chosen, depending on the priors. In 2-TMPL fits, the solution would be
directly dictated by the pre-selected $T_{\rm C}$ values, which might also
fall completely outside the $\chi^2=1$ contour.

\begin{figure*}
\centering
\includegraphics[width=12cm]{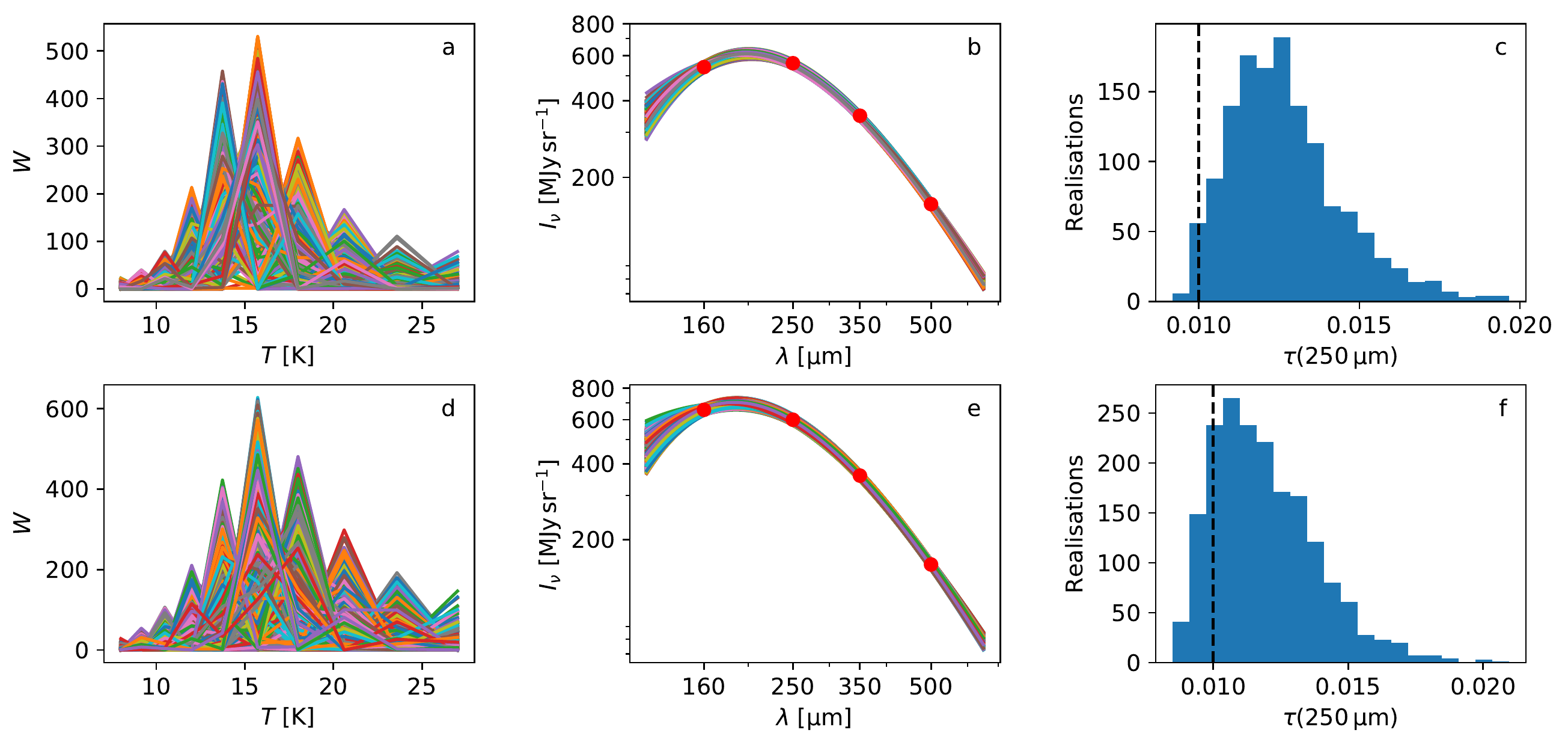}
\caption{
Fits to simulated observations with a 10-TMPL model. The observations consist
of 160, 250, 350, and 500\,$\mu$m intensities for a single $T=15$\,K MBB
function (upper frames) or a sum of 13\,K and 17\,K components of equal
optical depth (red circles in frames b and e). Frames a and d show random
realisations of the combinations of weights $W$ that result in less than 5\%
errors for all four predicted intensities (curves in frames b and e). The
frames c and f show the distributions of the $\tau$ values predicted by those
fits. The true optical depth is $\tau(250\mu{\rm m})=0.01$.
}
\label{fig:all_T_for_SED_A}
\end{figure*}

Based on the above, component degeneracies can be a problem already in a
two-component model. In multi-component models, these risks are only likely to
increase, and feasible solutions with $\chi^2\sim 1$ may be reached with many
parameter combinations that correspond to different optical depths. In
Fig.~\ref{fig:all_T_for_SED_A}, we simulate 160-500\,$\mu$m observations that
correspond to a single $T=15$\,K MBB spectrum or the sum of two MBB components
at 13\,K and 17\,K. We examined in this case the use of the 10-TMPL model
where the $T_{\rm C}$ values are set logarithmically between 8\,K to 27\,K. We
generated random weights for the ten components and registered the
combinations that fitted all SED measurements to better than 5\% accuracy.
Because this is a condition for the maximum error in any of the four bands, it
is more strict than the assumption of 3\% relative uncertainty that was used
in the previous tests. The detailed shapes of the $\tau/\tau_0$ distributions
(shown in frames c and f) depend on the way the simulation was done.
Nevertheless, they show that the recovered $\tau$ values are almost always
above the true values. The optical dept can be underestimated only in frame f,
when the fit relies on $T_{\rm C}$ components that fall between the true
temperatures of 13\,K and 17\,K. This is another illustration of the obvious
fact that four measurements cannot determine 12 parameters in a unique way.
Conversely, Fig.~\ref{fig:all_T_for_SED_A} can be interpreted to show how the
observed sources could have up to a factor of two differences in the optical
depths, in spite of showing observationally identical SEDs.

Strict priors can be set only if the nature of the object is well known. If
one uses a prior on temperature, such as $T\sim N(T_0, \delta T)$ in some
tests in this paper, the parameters $T_0$ and $\delta T$ will clearly be more
appropriate for some parts of the maps than for others. Even if $T_0$ is
varied spatially (e.g. following the temperature from the 1-MBB fit), the
parameter $\delta T$ still has a clear effect on the solution (a small $\delta
T$ leading to a lower $\tau$). As noted above, the significance of the model
errors means that the absolute accuracy of a column density measurement (match
to $\tau_0$) cannot be judged based on the $\chi^2$ values (match to observed
intensities).  Based on synthetic \Herschel 70-500\,$\mu$m observations of a
cloud simulation, \citet{Koepferl2017} also noted that fits with low $\chi^2$
values can correspond to very inaccurate estimates of temperature and column
density and, in their case, the fits could entirely fail close to hot sources.

The {\sc PPMAP} method \citep{Marsh2015} is another
multi-component SED fitting method that uses a combination of SED templates.
These can correspond different temperatures (typically a number larger than
the number of observed wavelengths) and optionally also to multiple $\beta$
values \citep[e.g.][]{Howard2019}. The model is thus in some respects similar
to $N$-TMPL and should be subject to some similar problems of parameter
degeneracy. The PPMAP model can use priors for example for the temperature.
Furthermore, the program looks for a solution that fits the observations to a
$\chi^2 \sim $1 accuracy with the smallest number of objects in the state
space. In a field like the IRDC model it might at each position use only a
couple of components that best match the local temperature distribution, also
avoiding excessive fitting of minor noise structures in the SEDs. On the other
hand, as discussed above, a low number of (SED) components is not a guarantee
of unbiased results and the same $\chi^2$ value may be reached with different
parameter combinations with different predictions of the optical depth. All
real source also have continuous temperature distributions, so the number of
components needed (and thus also the resulting $\tau$ estimates) will depend on
the observational noise.

\begin{figure*}
\centering
\includegraphics[width=12cm]{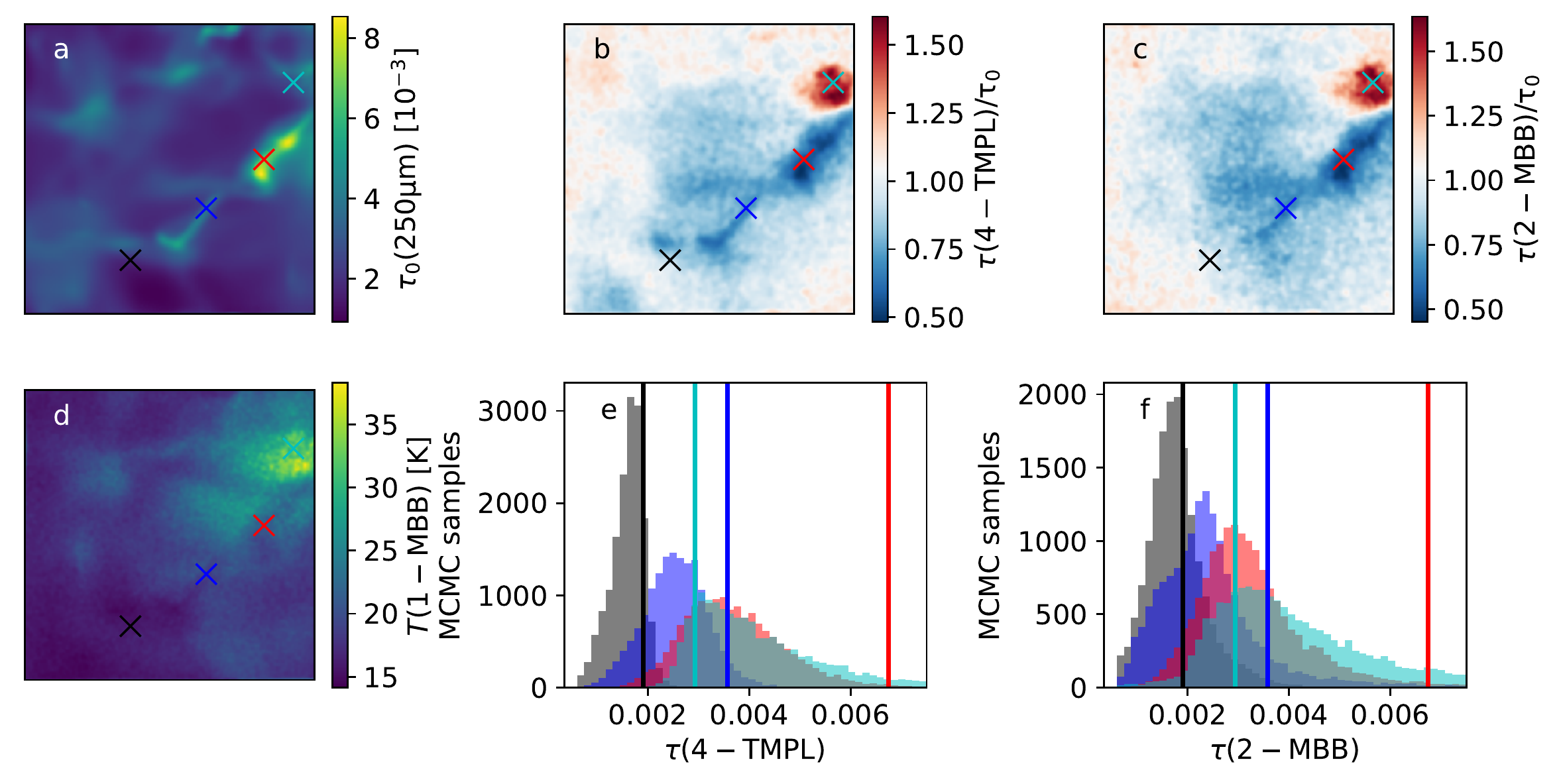}
\caption{
IRDC model observations fitted with MCMC methods. Frame a shows the true
$\tau(250\,\mu{\rm m})$ optical depths, and frame d the colour temperature
from the 1-MBB fit. Frames b and c show the $\tau$ maps from the 4-TMPL and
2-MBB fits (averages over the MCMC samples) divided by the true optical depth
$\tau_0$. Frames e and f show the posterior probability distributions for the
four positions marked in the b and c frames, and the true values $\tau_0$ are
indicated with vertical lines of the same colour (black, blue, red, and cyan).
}
\label{fig:TEST_IRDC_MCMC}
\end{figure*}

Figure~\ref{fig:TEST_IRDC_MCMC} shows examples of the $\tau$
distributions estimated with MCMC fits. The data are a subset of the IRDC
model maps, and the fitted models are 4-TMPL and 2-MBB with the same
parameters as in Sect.~\ref{results:IRDC}. All data are convolved to the same
resolution ($18\arcsec$ for an assumed $6\arcsec$ pixel size) to allow faster
calculations where each map pixel is fitted separately. These MCMC fits
(4-TMPL and 2-MBB models) of the $256 \times 256$ pixel maps took each less
than ten seconds (with 40000 MCMC steps after the burn-in phase). Therefore,
as long as convolution is not part of the model, MCMC remains feasible even
for the largest maps. For the assumed 3\% observational noise, the typical
uncertainty of an individual $\tau$ measurement is less than 50\%. However, in
those map positions, where the fits underestimate or overestimate the true
values the most, the true value is far in the tail of the probability
distribution. This is natural since even MCMC gives error estimates only
within the confines of the selected model. The uncertainties do not cover the
model errors, whether the selected model is appropriate or unique in
describing the given data.

In this paper, $\beta$ was mostly fixed to the value that was used in
generating the synthetic observations. If $\beta$ were included as a free
parameter in the SED fits, the formal uncertainties of the temperature
estimates would increase significantly due to the well-known anti-correlation
between the $T$ and $\beta$ parameters
\citep{Shetty2009b,Kelly2012,Juvela2013_TBmethods}. Low $\chi^2$ values would
be concentrated in curved regions in the ($T$, $\beta$) plane, where the
observational noise could cause large changes in the best-fit parameters. The
large uncertainty in $T$ will directly translate into large uncertainty in
$\tau$. The $\chi^2$ plane could even exhibit distinct minima that correspond
to significantly different temperatures and column density estimates
\citep{Juvela2012_bananasplit}. The $\chi^2$ values of the $N$-TMPL and
$N$-MBB fits could also exhibit multiple minima, even if $\beta$ we kept
constant.

\subsection{Correction to MBB fits}  \label{sect:correction}

Instead of relying on possibly ill-constrained multi-component models, one
possibility is to start with the 1-MBB fit and try to mitigate its bias. 
\citet{GCC-V} carried out radiative transfer (RT) modelling of dense clumps
and analysed the resulting synthetic surface brightness maps with the 1-MBB
method. The ratio between the true optical depth and the optical depth
estimated from the synthetic observations was then used to correct the 1-MBB
$\tau$ estimates. The accuracy of the correction depends on how well the RT
model matches the observed target, and this approach is therefore mainly
limited to sources with a simple structure. If the model matched observations
perfectly, it would directly provide good estimates for the column density,
without the need for any further SED analysis. However, it is more robust to
use RT models only to derive fractional corrections that are then applied to
1-MBB results. If the target is spherical (or otherwise geometrically simple),
methods like the Abell transformation \citep{Roy2014} provide a more
straightforward and computationally efficient way to estimate the temperature
and column density structure of a source. However, a full RT model still has
the advantage of presenting a physically self-consistent description of the
source, where the temperatures cannot vary in an arbitrary fashion but must be
consistent with the radiation sources, the density field, and the resulting
attenuation of the radiation field.  However, the increased complexity also
means that it is more difficult to find a radiative transfer model that would
be able to match all observations to the full precision of the observations
\citep{Juvela2013colden}.

One can ask, whether the observations themselves contain enough information to
go beyond the simplest 1-MBB fits. If the measurements do not sufficiently
constraint the solution, there will always be significant systematic errors
that depend on the construction of the SED models.

\begin{figure}
\centering
\includegraphics[width=9cm]{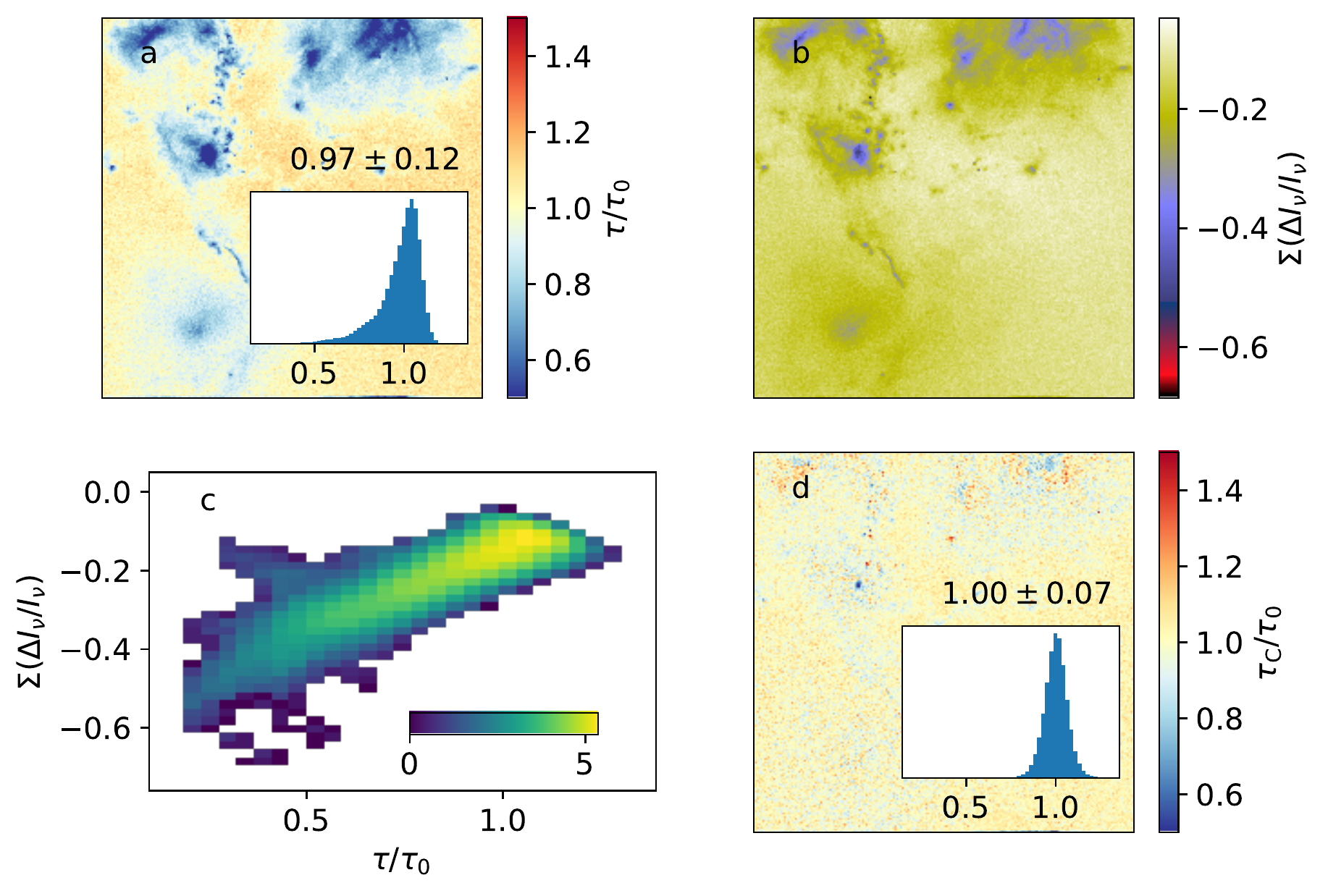}
\caption{
Empirical correction of the 1-MBB optical depth estimates of the IRDC cloud
model. Frame a shows the ratio $\tau/\tau_0$ between the original estimates
and the true optical depths. Frame b shows the quantity $\Sigma(\Delta
I_{\nu}/I_{\nu})$ that characterises the difference between each observed
spectrum and the 1-MBB fit. Frame c shows the correlation between this
quantity and the $\tau$ error, and Frame d shows the $\tau/\tau_0$ ratio
corrected for the trend in frame c. In frames a and d, the inserts show
histograms of the $\tau/\tau_0$ values (with linear y-axis). In frame c, the
colour scale corresponds to logarithmic point density.
}
\label{fig:corrected_MBB}
\end{figure}

The IRDC cloud of Sect.~\ref{results:IRDC} is the most complex model examined
in this paper, containing large temperature and density gradients.
Figure~\ref{fig:corrected_MBB}a shows again the 1-MBB results where the $\tau$
estimates are in many regions only 50\% of the correct value and the maximum
errors are close to a factor of five. The main effect of temperature
variations is to make the SED broader. With relative differences between the
observed intensities and the 1-MBB model fit, $r(\lambda)=(I_{\nu}^{\rm
1-MBB}-I_{\nu}^{\rm observed})/I_{\nu}^{\rm observed}$, we define a parameter
\begin{equation}
\Sigma(\Delta I_{\nu}/I_{\nu}) = 
r(160\,\mu{\rm m})-r(250\,\mu{\rm m})-r(350\,\mu{\rm m})+r(500\,\mu{\rm m}),
\label{eq:emp}
\end{equation}
which becomes more negative for wider SEDs. Figure~\ref{fig:corrected_MBB}b-c
shows that $\Sigma(\Delta I_{\nu}/I_{\nu})$ is well correlated with the bias
of the 1-MBB $\tau$ estimates. In Fig.~\ref{fig:corrected_MBB}d, the $\tau$
estimates have been corrected for this trend, using a third order polynomial
fit to the data in Fig.~\ref{fig:corrected_MBB}c. This simple empirical
correction is successful in eliminating most of the bias, although there is
only small improvement towards hot sources. The correction is also more
sensitive to noise (being a second-order correction to the 1-MBB fit), which
reduces the improvement in the accuracy of individual measurements (as 
suggested by the dispersion in Fig.~\ref{fig:corrected_MBB}c). The example
shows that the observations do contain more information than used of by the
1-MBB fit alone, but, with the assumed 3\% observational noise, there also are
limits to the potential improvement. 

We tested also the use of more general machine learning methods to derive the
corrections. A 246$\times$246 pixel part of the IRDC maps was used as the
training set for a small, four layers deep neural network. We omitted from the
maps only 5-pixel wide border regions that are affected by some edge effects
from the beam convolution. The training could be accomplished in a time that
is comparable to the previous direct multi-component SED fits, and the trained
network can then be used to quickly correct the 1-MBB estimates.

\begin{figure}
\centering
\includegraphics[width=9cm]{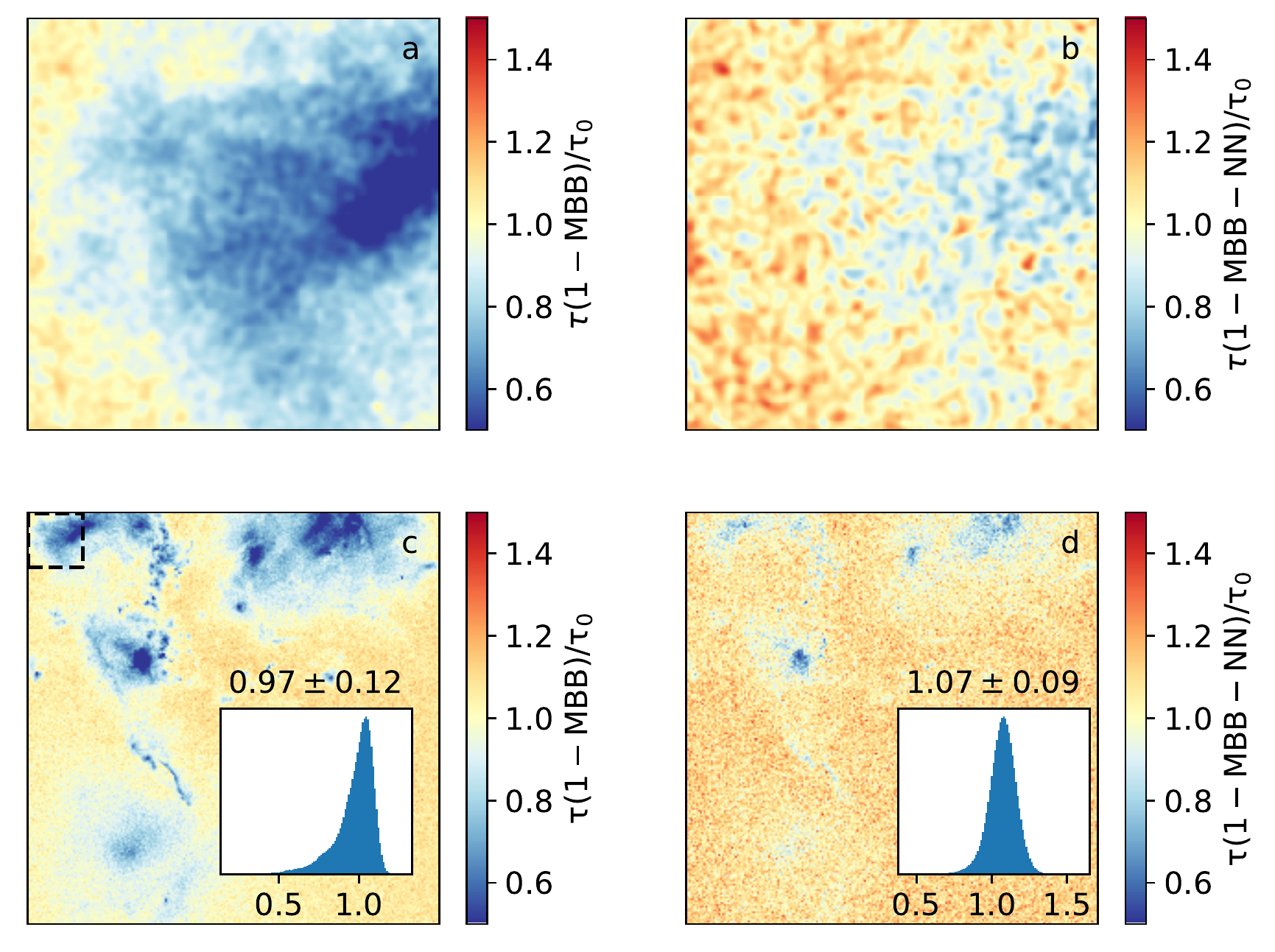}
\caption{
Correction to 1-MBB $\tau$ estimates calculated with neural nets. Frame a
shows the 246$\times$246 pixel map that is used as the training set. Frame $c$
shows the 1-MBB results for the whole IRDC map, where the location of the
training map is indicated with a dashed box in the upper left corner. Frames b
and d shows the corrected $\tau$ maps for the training set and for the full
1875$\times$1875 pixel map, respectively. The lower frames include histograms
of the $\tau/\tau_0$ values with their mean and standard deviation values.
}
\label{fig:plot_NN_comparison}
\end{figure}

Figure~\ref{fig:plot_NN_comparison} compares the accuracy between the original
1-MBB estimates and the neural-net corrected 1-MBB-NN estimates. For the
246$\times$246 pixel training set, the correction naturally works well.
However, it is not perfect, since Fig.~\ref{fig:plot_NN_comparison}b still
shows faintly the pattern that dominated the errors in the original 1-MBB
fits. The trained network was then used to correct data for the full 
1875$\times$1875 pixel map (Fig.~\ref{fig:plot_NN_comparison}c-d). The result
is rather similar to the empirical correction in
Fig.~\ref{fig:corrected_MBB}d. While in Fig.~\ref{fig:corrected_MBB} the
correction was estimated from the same map where it was then applied, in
Fig.~\ref{fig:plot_NN_comparison} the correction was derived from a very small
subset ($\sim 1.5$\% of the pixels, the rectangle marked in
Fig.~\ref{fig:plot_NN_comparison}c). The corrected $\tau$ values have a
symmetric error distribution. Although there is some positive bias
($\tau/\tau_0 \sim 1.07 \pm 0.09$), the maximum errors have been significantly
reduced. This is encouraging regarding the possibility of finding useful
corrections, possibly tuned for a given type of sources, that are general
enough to be applied also to new observations.

\subsection{Computational cost} \label{sect:calculations}

The computational cost of the different SED models varied by about four orders
of magnitude. Once the data are convolved to the resolution of the longest
wavelength observations, the 1-MBB fits of the 1875$\times$1875 maps could be
fitted in just a couple of seconds (Appendix~\ref{app:MBB}). Even the 
empirical corrections of Sect.~\ref{sect:correction} would add little to the
run times. This can provide a more robust way to estimate optical depths,
considering the potential degeneracies in the multi-component fits, although
the angular resolution is limited to the lowest common resolution.

When the beam convolution is part of the fit, the run times increase by orders
of magnitude. For the $N$-TMPL models only, the optimisation can be done
faster in Fourier space (Appendix~\ref{app:MBB}). In the case of IRDC maps
(1875$\times$1875 pixels and four wavelengths), the run times were of the
order of one minute instead of the order of one hour for the fits optimised in
the real space. The run times also depend on the chosen convergence criteria.
The $\chi^2$ values tended to decrease initially fast, with only one or two
percent further improvement over a larger number of further iterations.

The beam convolution is a major part of the computational cost. For small maps
of up to $\sim 256 \times 256$ pixels, the use of a GPU for the FFT
computations provides no benefits, but for the largest maps (1875$\times$1875
pixels) the speed-up was a factor of a few. The use of FFTs can cause
artefacts at the map boundaries, due to the assumed data periodicity. It is
possible to carry out the convolutions in real space, by direct summation of
the pixel values over the beams. This gives more freedom in the use of
measurement uncertainties (including missing values and potentially even error
correlations) and in the handling of the map edges. This remains a viable
option for smaller maps, especially when GPUs are used.

The MCMC fitting is typically more time consuming than direct $\chi^2$
minimisation. On the other hand, it could provide more complete information of
the uncertainties than just using the local $\chi^2$ gradients. Some examples
of the use of MCMC calculations were given in
Figs.~\ref{fig:Figure1}-\ref{fig:Figure3} and in
Sect.~\ref{sect:degeneracies}, where each map pixel was still fitted
separately. However, MCMC turns out to be viable for maps of moderate size
even when the beam convolution is part of the model optimisation.
Appendix~\ref{app:MCMC} shows one example, where $320 \times 320$ pixel maps
are fitted with the 3-TMPL model, the MCMC runs taking about ten minutes on a
GPU. The MCMC error estimates are consistent with the observed scatter in the
$\tau$ estimates. However, we emphasise again that these formal error
estimates, whether based on the $\chi^2$ values or MCMC sampling, do not
account for the systematic errors that were usually the dominant error
component in our tests.

\subsection{Further improvements}

All fits in Sect.~\ref{sect:results} were based on SED models that consist of
discrete temperature components. This is not a fundamental limitation, and the
basis functions could equally be based on parameterised temperature
distributions. Appendix~\ref{app:MCMC2} shows one example, where the SED
templates correspond to different Gaussian distributions of the optical depth
versus dust temperature. The model parameters are thus the mean value $\langle
T \rangle$ and the width $\delta T$ of the temperature distribution and the
optical depth. Although the run time was longer (about one hour), the results
were in this case also more accurate than those of the 3-TMPL model
(Sect.~\ref{app:MCMC}) that has the same number of free parameters.

The model used in Appendix~\ref{app:MCMC2} is different from both the $N$-TMPL
and $N$-MBB ones, and it combines a low number of free parameters with the
flexibility of adjusting the mean temperature in a continuous fashion. The
MCMC runs used a lookup table of pre-calculated dust spectra. The same
approach could be used for any SEDs, which could be based on empirical (e.g.
log-normal or broken power laws) or model-based temperature distributions
\citep[e.g.][]{Dale2001, Galliano2011, Galliano2018}. For the former,
\citet{Desert2022} advocated the use the logarithm of temperature as the more
natural parameter. The basis functions could also directly encode prior
information, such as the avoidance of unphysically low temperatures.

In this paper we have not considered the possible correlations between the
input parameters. If the measurements of the different bands are correlated,
the $\chi^2$ values are calculated by replacing the sum of $(X_i-\hat
X_i)^2/\sigma(X_i)^2$ values with $(X-\hat X) \Sigma^{-1} (X-\hat X)^{T}$,
where $X$ are the measured values, $\hat X$ the model prediction, and $\Sigma$
the covariance matrix of the observations. With a handful of frequency bands,
the effect on the run times would be at most a factor of a few. The spatial
correlations of the input data have no effect on the 1-MBB fits, which are
done independently for each resolution element. The correlations should
nevertheless be taken into account when the data are initially convolved to a
common resolution. If the convolution is part of the fitting procedure, the
handling of the spatial correlations would have a large impact on the run
times. Depending on the beam sizes, each pixel might have non-zero covariances
with tens or even hundreds of other pixels. The additional cost would come
from the increased complexity in the convolution (to be done now in real
space) and in the evaluation of the goodness-of-fit function. The
implementation would be more straightforward in the MCMC framework, where the
derivatives of the probability function are not needed. However, there would
still be significant practical limitations on the size of the maps that could
be analysed.

\section{Conclusions}  \label{sect:conclusions}

We have examined the analysis of dust emission spectra with SED models that
consist of $N$-fixed SED templates (models $N$-TMPL, $N>1$) or $N$ MBB
functions (models $N$-MBB, $N\ge 1$). Tests with synthetic observations have
led to the following conclusions.

\begin{itemize}
\item The $N$-TMPL and $N$-MBB models can improve the
standard single-component 1-MBB fits, but they require a careful model setup and, in
the case of large $N$, the use of appropriate priors.
\item The accuracy of the $N$-TMPL models critically depends on the selection 
of its SED templates. Errors increase rapidly when the temperature of the
emission falls outside the temperature range covered by the templates.
\item The $N$-MBB models adjust better to large temperature variations in the
source maps, but they may need to be combined with priors to avoid very low
temperatures that would cause large noise in the $\tau$ estimates. This is
true especially if there are many pixels per beam and if the SED of individual
pixels is thus not well constrained. 
\item The fitting of multi-component models is computationally feasible even
for large maps, provided that the gradients of the $\chi^2$ function are
calculated analytically. In our tests, SED fits were performed for maps of up
to $1875 \times 1875$ pixels.
\item When the beam convolution is part of every optimisation step, the use of 
GPUs can significantly reduce the run times.
\item We have presented alternative $N$-TMPL fits that were based on the
optimisation in the Fourier space. This results in significant speed-up in the
computations, but with some limitations in the use of priors and general error
estimates.
\item MCMC remains feasible for maps of at least up to $\sim 500\times 500$ pixels
(run times of approximately one hour or less), even when each MCMC step includes the
convolution of the model-predicted maps. MCMC provides better estimates of the
statistical errors, but these do not give any indication of the often larger
systematic errors.
\item Multi-component fits can contain fundamental degeneracies, where
(in the studied cases) the same observed intensities could correspond to a
factor of two differences in the true optical depths. 
\item The bias of the 1-MBB fits can be decreased based on observed data alone
by analysing the widening of the SEDs. Neural networks provide a general way
to derive such corrections.  
\end{itemize}

The results call for caution when interpreting the results from
multi-component SED fits. This is particularly true if the number of free
parameters exceeds the number of direct observational constraints and the
results thus clearly depend on the chosen priors. Having low $\chi^2$ values for the
fits does not mean that the corresponding $\tau$ estimates would be correct,
but the comparison of alternative SED models can give some indication of the
magnitude of the model errors. Our initial results suggest that models with
parameterised continuous temperature distributions might perform better than
the models (such as $N$-TMPL and $N$-MBB) where only discrete temperatures are
used. This needs to be investigated further in a wider range of test cases.
Radiative transfer modelling also remains a possibility in the analysis of
emission from simple sources, and it is not much more time consuming than the
fitting of a complex multi-component SED model.

\begin{acknowledgements}
This work was supported by the Academy of Finland grant No. 348342.
\end{acknowledgements}

\bibliographystyle{aa}
\bibliography{my.bib}

\begin{thebibliography}{43}
\expandafter\ifx\csname natexlab\endcsname\relax\def\natexlab#1{#1}\fi

\bibitem[{{Andrae}(2010)}]{Andrae2010a}
{Andrae}, R. 2010, arXiv e-prints, arXiv:1009.2755

\bibitem[{{Bonnor}(1956)}]{Bonnor1956}
{Bonnor}, W.~B. 1956, MNRAS, 116, 351

\bibitem[{{Chang} {et~al.}(2021){Chang}, {Zhou}, {Lamperti}, {Saintongel},
  {Esimbek}, {Li}, {He}, {Qiu}, {Li}, {Zhou}, {Tang}, {Wu}, {Ji}, {Zhao}, \&
  {Zhou}}]{Chang2021}
{Chang}, Z., {Zhou}, J., {Lamperti}, I., {et~al.} 2021, \apj, 915, 51

\bibitem[{{Chastenet} {et~al.}(2021){Chastenet}, {Sandstrom}, {Chiang},
  {Hensley}, {Draine}, {Gordon}, {Koch}, {Leroy}, {Utomo}, \&
  {Williams}}]{Chastenet2021}
{Chastenet}, J., {Sandstrom}, K., {Chiang}, I.-D., {et~al.} 2021, \apj, 912,
  103

\bibitem[{{Compi{\`e}gne} {et~al.}(2011){Compi{\`e}gne}, {Verstraete}, {Jones},
  {Bernard}, {Boulanger}, {Flagey}, {Le Bourlot}, {Paradis}, \&
  {Ysard}}]{Compiegne2011}
{Compi{\`e}gne}, M., {Verstraete}, L., {Jones}, A., {et~al.} 2011, \aap, 525,
  A103+

\bibitem[{{Crapsi} {et~al.}(2007){Crapsi}, {Caselli}, {Walmsley}, \&
  {Tafalla}}]{Crapsi2007}
{Crapsi}, A., {Caselli}, P., {Walmsley}, M.~C., \& {Tafalla}, M. 2007, \aap,
  470, 221

\bibitem[{{Dale} {et~al.}(2001){Dale}, {Helou}, {Contursi}, {Silbermann}, \&
  {Kolhatkar}}]{Dale2001}
{Dale}, D.~A., {Helou}, G., {Contursi}, A., {Silbermann}, N.~A., \&
  {Kolhatkar}, S. 2001, \apj, 549, 215

\bibitem[{{D{\'e}sert}(2022)}]{Desert2022}
{D{\'e}sert}, F.-X. 2022, \aap, 659, A70

\bibitem[{{Draine}(2003)}]{Draine2003ARAA}
{Draine}, B.~T. 2003, \araa, 41, 241

\bibitem[{{Ebert}(1955)}]{Ebert1955}
{Ebert}, R. 1955, Zeitschrift fur Astrophysik, 37, 217

\bibitem[{{Galametz} {et~al.}(2012){Galametz}, {Kennicutt}, {Albrecht},
  {Aniano}, {Armus}, {Bertoldi}, {Calzetti}, {Crocker}, {Croxall}, {Dale},
  {Donovan Meyer}, {Draine}, {Engelbracht}, {Hinz}, {Roussel}, {Skibba},
  {Tabatabaei}, {Walter}, {Weiss}, {Wilson}, \& {Wolfire}}]{Galametz2012}
{Galametz}, M., {Kennicutt}, R.~C., {Albrecht}, M., {et~al.} 2012, \mnras, 425,
  763

\bibitem[{{Galliano}(2018)}]{Galliano2018}
{Galliano}, F. 2018, \mnras, 476, 1445

\bibitem[{{Galliano} {et~al.}(2011){Galliano}, {Hony}, {Bernard}, {Bot},
  {Madden}, {Roman-Duval}, {Galametz}, {Li}, {Meixner}, {Engelbracht},
  {Lebouteiller}, {Misselt}, {Montiel}, {Panuzzo}, {Reach}, \&
  {Skibba}}]{Galliano2011}
{Galliano}, F., {Hony}, S., {Bernard}, J.-P., {et~al.} 2011, \aap, 536, A88

\bibitem[{{Gordon} {et~al.}(2014){Gordon}, {Roman-Duval}, {Bot}, {Meixner},
  {Babler}, {Bernard}, {Bolatto}, {Boyer}, {Clayton}, {Engelbracht}, {Fukui},
  {Galametz}, {Galliano}, {Hony}, {Hughes}, {Indebetouw}, {Israel}, {Jameson},
  {Kawamura}, {Lebouteiller}, {Li}, {Madden}, {Matsuura}, {Misselt}, {Montiel},
  {Okumura}, {Onishi}, {Panuzzo}, {Paradis}, {Rubio}, {Sandstrom}, {Sauvage},
  {Seale}, {Sewi{\l}o}, {Tchernyshyov}, \& {Skibba}}]{Gordon2014}
{Gordon}, K.~D., {Roman-Duval}, J., {Bot}, C., {et~al.} 2014, \apj, 797, 85

\bibitem[{{Harju} {et~al.}(2008){Harju}, {Juvela}, {Schlemmer}, {Haikala},
  {Lehtinen}, \& {Mattila}}]{Harju2008}
{Harju}, J., {Juvela}, M., {Schlemmer}, S., {et~al.} 2008, \aap, 482, 535

\bibitem[{{Haugb{\o}lle} {et~al.}(2018){Haugb{\o}lle}, {Padoan}, \&
  {Nordlund}}]{Haugbolle2018}
{Haugb{\o}lle}, T., {Padoan}, P., \& {Nordlund}, {\r{A}}. 2018, \apj, 854, 35

\bibitem[{{Howard} {et~al.}(2019){Howard}, {Whitworth}, {Marsh}, {Clarke},
  {Griffin}, {Smith}, \& {Lomax}}]{Howard2019}
{Howard}, A.~D.~P., {Whitworth}, A.~P., {Marsh}, K.~A., {et~al.} 2019, \mnras,
  489, 962

\bibitem[{{J{\o}rgensen} {et~al.}(2020){J{\o}rgensen}, {Belloche}, \&
  {Garrod}}]{Jorgensen2020}
{J{\o}rgensen}, J.~K., {Belloche}, A., \& {Garrod}, R.~T. 2020, \araa, 58, 727

\bibitem[{{Juvela} {et~al.}(2015{\natexlab{a}}){Juvela}, {Demyk}, {Doi},
  {Hughes}, {Lef{\`e}vre}, {Marshall}, {Meny}, {Montillaud}, {Pagani},
  {Paradis}, {Ristorcelli}, {Malinen}, {Montier}, {Paladini}, {Pelkonen}, \&
  {Rivera-Ingraham}}]{GCC-VI}
{Juvela}, M., {Demyk}, K., {Doi}, Y., {et~al.} 2015{\natexlab{a}}, \aap, 584,
  A94

\bibitem[{{Juvela} {et~al.}(2013{\natexlab{a}}){Juvela}, {Malinen}, \&
  {Lunttila}}]{Juvela2013colden}
{Juvela}, M., {Malinen}, J., \& {Lunttila}, T. 2013{\natexlab{a}}, \aap, 553,
  A113

\bibitem[{{Juvela} \& {Mannfors}(2022)}]{Juvela2022b}
{Juvela}, M. \& {Mannfors}, E. 2022, \aap, submitted

\bibitem[{{Juvela} {et~al.}(2022){Juvela}, {Mannfors}, {Liu}, \&
  {Toth}}]{Juvela2022a}
{Juvela}, M., {Mannfors}, E., {Liu}, T., \& {Toth}, L.~V. 2022, arXiv e-prints,
  arXiv:2208.01894

\bibitem[{{Juvela} {et~al.}(2013{\natexlab{b}}){Juvela}, {Montillaud}, {Ysard},
  \& {Lunttila}}]{Juvela2013_TBmethods}
{Juvela}, M., {Montillaud}, J., {Ysard}, N., \& {Lunttila}, T.
  2013{\natexlab{b}}, \aap, 556, A63

\bibitem[{{Juvela} {et~al.}(2015{\natexlab{b}}){Juvela}, {Ristorcelli},
  {Marshall}, {Montillaud}, {Pelkonen}, {Ysard}, {McGehee}, {Paladini},
  {Pagani}, {Malinen}, {Rivera-Ingraham}, {Lef{\`e}vre}, {T{\'o}th}, {Montier},
  {Bernard}, \& {Martin}}]{GCC-V}
{Juvela}, M., {Ristorcelli}, I., {Marshall}, D.~J., {et~al.}
  2015{\natexlab{b}}, \aap, 584, A93

\bibitem[{{Juvela} \& {Ysard}(2012{\natexlab{a}})}]{Juvela2012_bananasplit}
{Juvela}, M. \& {Ysard}, N. 2012{\natexlab{a}}, \aap, 541, A33

\bibitem[{{Juvela} \& {Ysard}(2012{\natexlab{b}})}]{Juvela2012_Tmix}
{Juvela}, M. \& {Ysard}, N. 2012{\natexlab{b}}, \aap, 539, A71

\bibitem[{{Kelly} {et~al.}(2012){Kelly}, {Shetty}, {Stutz}, {Kauffmann},
  {Goodman}, \& {Launhardt}}]{Kelly2012}
{Kelly}, B.~C., {Shetty}, R., {Stutz}, A.~M., {et~al.} 2012, \apj, 752, 55

\bibitem[{{Koepferl} {et~al.}(2017){Koepferl}, {Robitaille}, \&
  {Dale}}]{Koepferl2017}
{Koepferl}, C.~M., {Robitaille}, T.~P., \& {Dale}, J.~E. 2017, \apj, 849, 1

\bibitem[{{K{\"o}hler} {et~al.}(2015){K{\"o}hler}, {Ysard}, \&
  {Jones}}]{Kohler2015}
{K{\"o}hler}, M., {Ysard}, N., \& {Jones}, A.~P. 2015, \aap, 579, A15

\bibitem[{{Lamperti} {et~al.}(2019){Lamperti}, {Saintonge}, {De Looze},
  {Accurso}, {Clark}, {Smith}, {Wilson}, {Brinks}, {Brown}, {Bureau},
  {Clements}, {Eales}, {Glass}, {Hwang}, {Lee}, {Lin}, {Michalowski},
  {Sargent}, {Williams}, {Xiao}, \& {Yang}}]{Lamperti2019}
{Lamperti}, I., {Saintonge}, A., {De Looze}, I., {et~al.} 2019, \mnras, 489,
  4389

\bibitem[{{Malinen} {et~al.}(2011){Malinen}, {Juvela}, {Collins}, {Lunttila},
  \& {Padoan}}]{Malinen2011}
{Malinen}, J., {Juvela}, M., {Collins}, D.~C., {Lunttila}, T., \& {Padoan}, P.
  2011, \aap, 530, A101+

\bibitem[{{Marsh} {et~al.}(2015){Marsh}, {Whitworth}, \& {Lomax}}]{Marsh2015}
{Marsh}, K.~A., {Whitworth}, A.~P., \& {Lomax}, O. 2015, \mnras, 454, 4282

\bibitem[{{Marsh} {et~al.}(2017){Marsh}, {Whitworth}, {Lomax}, {Ragan},
  {Becciani}, {Cambr{\'e}sy}, {Di Giorgio}, {Eden}, {Elia}, {Kacsuk},
  {Molinari}, {Palmeirim}, {Pezzuto}, {Schneider}, {Sciacca}, \&
  {Vitello}}]{Marsh2017}
{Marsh}, K.~A., {Whitworth}, A.~P., {Lomax}, O., {et~al.} 2017, \mnras, 471,
  2730

\bibitem[{{Mathis} {et~al.}(1983){Mathis}, {Mezger}, \& {Panagia}}]{Mathis1983}
{Mathis}, J.~S., {Mezger}, P.~G., \& {Panagia}, N. 1983, \aap, 128, 212

\bibitem[{{Motte} {et~al.}(2018){Motte}, {Bontemps}, \& {Louvet}}]{Motte2018}
{Motte}, F., {Bontemps}, S., \& {Louvet}, F. 2018, \araa, 56, 41

\bibitem[{{Pagani} {et~al.}(2015){Pagani}, {Lef{\`e}vre}, {Juvela}, {Pelkonen},
  \& {Schuller}}]{Pagani2015}
{Pagani}, L., {Lef{\`e}vre}, C., {Juvela}, M., {Pelkonen}, V.-M., \&
  {Schuller}, F. 2015, \aap, 574, L5

\bibitem[{{Paradis} {et~al.}(2012){Paradis}, {Paladini}, {Noriega-Crespo},
  {M{\'e}ny}, {Piacentini}, {Thompson}, {Marshall}, {Veneziani}, {Bernard}, \&
  {Molinari}}]{Paradis2012_500um}
{Paradis}, D., {Paladini}, R., {Noriega-Crespo}, A., {et~al.} 2012, \aap, 537,
  A113

\bibitem[{{Paradis} {et~al.}(2010){Paradis}, {Veneziani}, {Noriega-Crespo},
  {Paladini}, {Piacentini}, {Bernard}, {de Bernardis}, {Calzoletti},
  {Faustini}, {Martin}, {Masi}, {Montier}, {Natoli}, {Ristorcelli}, {Thompson},
  {Traficante}, \& {Molinari}}]{Paradis2010}
{Paradis}, D., {Veneziani}, M., {Noriega-Crespo}, A., {et~al.} 2010, \aap, 520,
  L8

\bibitem[{{Planck Collaboration}(2014{\natexlab{a}})}]{planck2013-p06b}
{Planck Collaboration}. 2014{\natexlab{a}}, \aap, 571, A11

\bibitem[{{Planck Collaboration}(2014{\natexlab{b}})}]{planck2013-XVII}
{Planck Collaboration}. 2014{\natexlab{b}}, \aap, 566, A55

\bibitem[{{Roy} {et~al.}(2014){Roy}, {Andr{\'e}}, {Palmeirim}, {Attard},
  {K{\"o}nyves}, {Schneider}, {Peretto}, {Men'shchikov}, {Ward-Thompson},
  {Kirk}, {Griffin}, {Marsh}, {Abergel}, {Arzoumanian}, {Benedettini}, {Hill},
  {Motte}, {Nguyen Luong}, {Pezzuto}, {Rivera-Ingraham}, {Roussel}, {Rygl},
  {Spinoglio}, {Stamatellos}, \& {White}}]{Roy2014}
{Roy}, A., {Andr{\'e}}, P., {Palmeirim}, P., {et~al.} 2014, \aap, 562, A138

\bibitem[{{Shetty} {et~al.}(2009{\natexlab{a}}){Shetty}, {Kauffmann}, {Schnee},
  \& {Goodman}}]{Shetty2009b}
{Shetty}, R., {Kauffmann}, J., {Schnee}, S., \& {Goodman}, A.~A.
  2009{\natexlab{a}}, \apj, 696, 676

\bibitem[{{Shetty} {et~al.}(2009{\natexlab{b}}){Shetty}, {Kauffmann}, {Schnee},
  {Goodman}, \& {Ercolano}}]{Shetty2009a}
{Shetty}, R., {Kauffmann}, J., {Schnee}, S., {Goodman}, A.~A., \& {Ercolano},
  B. 2009{\natexlab{b}}, \apj, 696, 2234

\end{thebibliography}

\appendix

\section{Fitting algorithms} \label{app:MBB}

When the beam-convolved model predictions are compared to observations, the
fitting is accomplished by minimising the function
\begin{equation}
\chi^2 = 
    \sum_{i,k} \frac{S_{i,k} - \sum_c M_{i,c} \otimes B_k}{ \delta S_{i,k}^2} 
=   \sum_{i,k} \frac{S_{i,k} - {\hat S}_{i,k}        }{ \delta S_{i,k}^2},
\label{eq:0001}
\end{equation}
where $i$ refers to the map position and $k$ to the frequency band, and the
model is the sum of multiple components indexed with $c$. The symbol
`$\otimes$' stands for the convolution between a model map $M_{i,c}$ and the
telescope beam $B_k$, and ${\hat S}_{k}$ denotes the convolved model
predictions. The summation goes over all map positions $i$ and bands $k$. 

The functions $M_{i,c}$ are sums of SED components. The $N$-MBB fits use sums
of MBB functions
\begin{equation}
M_{i,c,k} =   MBB(I_{i,c}, T_{i,c}, \beta_{i,c})_{i,k},
\end{equation}
where $I_{i,c}$ is the intensity at a reference wavelength (see below), and
the result is also a function of the band $k$. In this paper, $\beta$ was
kept constant and only $I_{i,c}$ and $T_{i,c}$ are considered free
parameters. The $N$-TMPL model is somewhat simpler,
\begin{equation}
M_{i,c,k} =  W_{i,c} C_{c,k},
\end{equation}
with free parameters $W_{i,c,k}$ and the fixed SED templates $C_{c,k}$.

When the $\chi^2$ functions is minimised with the conjugate-gradient method,
only the function values and the first derivatives with respect to the
optimised parameters $X_{i,c}$ are needed. The derivatives with respect to
$X_{i_0,j_0}$ can be written as
\begin{equation}
\frac{\partial \chi_k^2}{\partial X_{i_0, c_0}}  =
 - 2  \sum_{k}
 \left[
 \left( \frac{S_{i,k}-{\hat S} _{i,k}}{dS_{i,k}}  \right)^2  \otimes B_{i,k}
 \right]
 \frac{\partial M_{i,c,k}}{\partial X_{i_0,c_0}}.
 \label{eq:c}
\end{equation}
Since the model prediction $\hat{S}$ itself is already the result of
convolution, the gradient calculation thus includes two convolution steps.
This can be shown by direct derivation of Eq.~(\ref{eq:0001}), by replacing
the convolution there with an explicit summation over the beam. 

The last term in Eq.~(\ref{eq:c}) is the derivative of the model with respect
to one of its parameter. In the sum, this is reduced to a single term (pixel
$i_0$ and component $c_0$).
If $M$ is the MBB function, it may be numerically beneficial to express it
using intensities relative to the value at a suitable reference frequency
$\nu_0$,
\begin{equation}
M(I, T, \beta)_{i,c,k} =  
I_{i,c} \,\, 
\frac{ \exp(h \nu_0/(kT_{i,c}))-1}{\exp(h \nu /(kT_{i,c}))-1},
\end{equation}
where frequency $\nu_0$ is the reference frequency and $\nu$ the frequency
of the band $k$. The full $M_{i,k}$ function is the sum of the individual
components $M_{i,c,k}$. Using the values of the functions $M_{i,c}$ itself, the
final partial derivatives in Eq.~(\ref{eq:c}) can be written as
\begin{equation}
\frac{\partial M_{i,c,k}}{\partial I_{i,c}} =  M_{i,k}/I_{i,c},
\label{eq:d}
\end{equation}
\begin{equation}
\frac{\partial M_{i,c,k}}{\partial T_{i,c}} =  \frac{M_{i,k} h \nu}{T_{i,c}^2}
\left( 
\frac{\nu  }{1-\exp(-h \nu   / (k T_{i,c}))} - 
\frac{\nu_0}{1-\exp(-h \nu_0 / (k T_{i,c}))}
\right)
\end{equation}
and
\begin{equation}
\frac{\partial M_{i,c,k}}{\partial \beta_{i,c}} =  M_{i,c,k} \log(\nu/\nu_0),
\end{equation}
thus replacing $X_{i_0,c_0}$ with $I$, $T$, or $\beta$. 
In the case of the $N$-TMPL fits, the derivatives $\partial M/\partial X$ are
directly elements of the SED templates
\begin{equation}
\frac{\partial M_{i,c,k}}{\partial W_{i,c}} = C_{i,k}.
\end{equation}

The non-negativity of the parameters $I$ could be enforced with an additional
penalty function or using an optimisation routine that directly support the
use of constraints. We chose to optimise the parameter $\alpha=\log(I)$, which
also guarantees the positivity of the intensity values, $I=e^\alpha$. For
$N$-MBB fits, Eq.~(\ref{eq:d}) becomes in this case
\begin{equation}
\frac{\partial M_{i,c}}{\partial \alpha_{i,c}} =  M_{i,c},
\end{equation}
where $M$ itself is of course evaluated by replacing $I$ with $\exp(\alpha)$.
In the $N$-TMPL fits, all $\partial M/\partial X$ derivatives are of the form
\begin{equation}
\frac{\partial M_{i,c,k}}{\partial W_{i,c}} = e^{W_{i,c}} C_{c,k},
\end{equation}
if the original weights $W_{i,c}$ are replaced with terms $e^{W_{i,c}}$.

When the optimisation of the $N$-TMPL model is done in Fourier space (models
$N$-TMPL-F), observations $S_{i,k}$ and the weight maps $W_{i,c}$ are replaced
in Eq.~(\ref{eq:0001}) by their Fourier counterparts and the convolution is
replaced by direct multiplication with the Fourier-transformed beams. This is
possible because the $N$-TMPL model is linear with respect to SED templates
and
\begin{equation}
S_k \approx \sum_c W_c C_{c,k}
\Rightarrow
{\cal F}(S_k) \approx \sum_c C_{c,k} {\cal F}(W_c),
\end{equation}
as used in Eq.~(\ref{eq:FOU}).

\begin{figure}
\centering
\includegraphics[width=9cm]{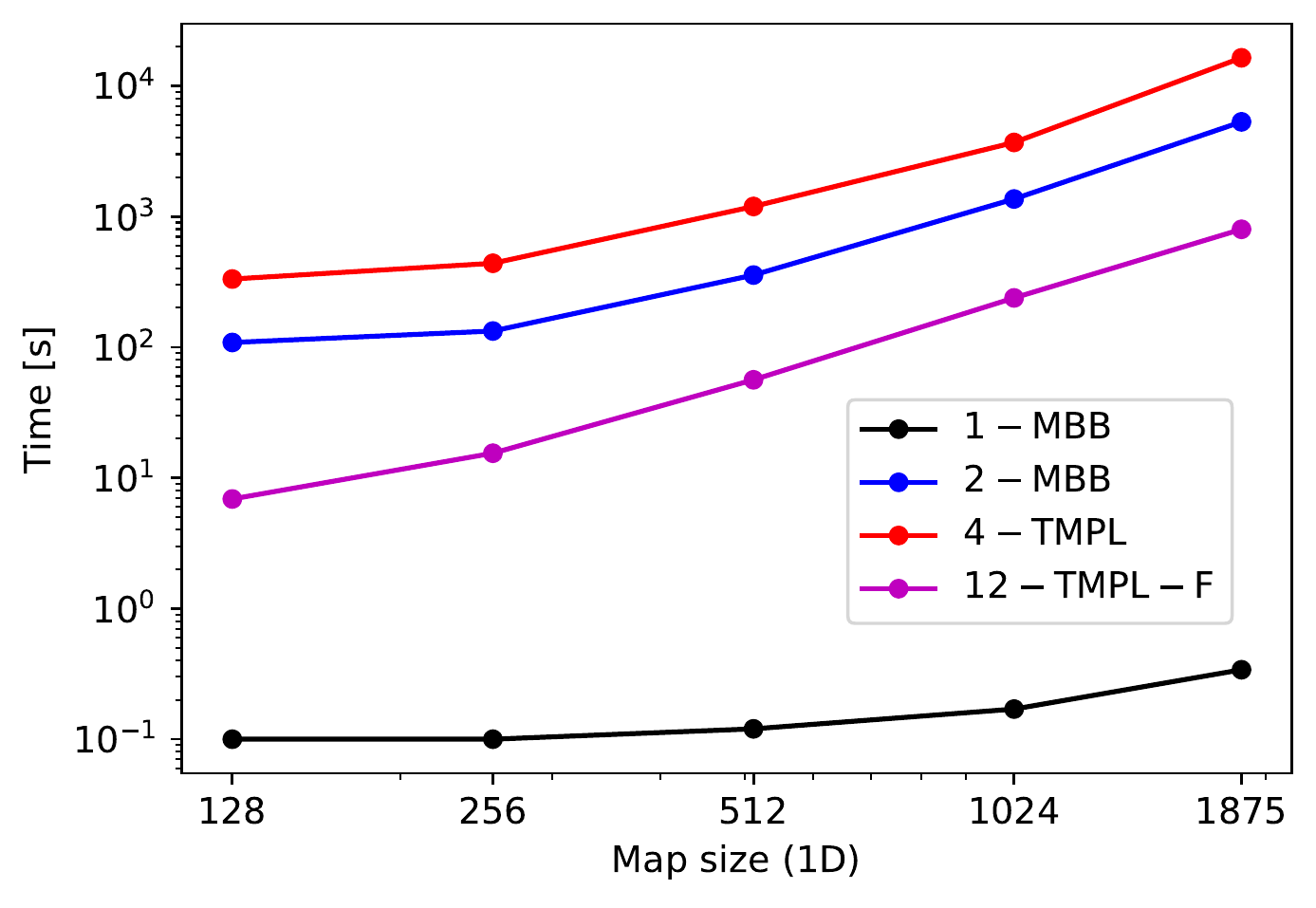}
\caption{
Comparison of run times of different SED fits. The observations are from the
IRDC model, and the map size is varied from $128\times 128$ to $1875 \times
1875$ pixels. Apart from 1-MBB, beam convolution is always part of model that
is being fitted. The number of fitted SED components is given in the legend,
and $TMPL-F$ stands for optimisation in the Fourier space.
}
\label{fig:timings}
\end{figure}

The relative run times depend on many factors that include the map size, the
number of fitted components, having convolutions as part of the model, and 
the computational methods used for the convolutions (i.e. either in real space
or via FFTs). Figure~\ref{fig:timings} compares the run time of the 1-MBB fit
to 2-MBB, 4-TMPL, and 12-TMPL-F models for maps of different sizes. The
observations are for the IRDC model and consist of 160, 250, 350, and
500\,$\mu$m maps. The 1-MBB fit is by far the fastest, because the
optimisation is done pixel by pixel and model does not include beam
convolution. In this case, all maps are first convolved to the same resolution
before the SED fitting. The quoted run times are for the optimisation only,
and these remain below one second even for the largest maps (when performed on
a GPU). The other methods are slower because of the direct additional cost of
the convolutions (although that cost is small in case of TMPL-F) and because
the convolution connects all parameters and makes the model fitting a more
difficult global optimisation problem. The 12-TMPL-F run times correspond to a
full run to a converged solution, where all calculations (including the
conjugate gradient optimiser) are run on a GPU. In the case of the 2-MBB and
4-TMPL fits, the run times correspond to 13000 function evaluations. This may
be sufficient in some cases but a good convergence may require even 2-3 times
longer runs. In the 2-MBB and 4-TMPL fits, the optimiser is run on the host
machine but the FFT convolutions are performed on a GPU. 

As mentioned in Sect.~\ref{sect:calculations}, FFT calculations were faster on
a GPU only for map sizes larger than 256$\times$256 pixels. Because the beam
sizes correspond to less than ten pixels (FWHM=36$\arcsec$ at 500\,$\mu$m,
compared to 6$\arcsec$ pixels), the convolutions could also be done with a GPU
in real space, without a drastic increase in the run times\footnote{\tt
https://www.interstellarmedium.org/numerical\_tools/ fits\_images}. The cost
of the real-space convolution also increases only directly proportionally to
the map dimensions ($\propto N \times M$) instead of the $N M \log(M N)$
scaling of 2D FFTs.

The calculations for Fig.~\ref{fig:timings} were performed on a laptop with a
discrete GPU (which is nevertheless three generations behind the most recent
hardware available). When the optimiser itself was run on the host computer,
the run times were mostly bound by the host (the single-threaded program)
rather than the FFT calculations on the GPU.

\section{Additional fits of the full IRDC maps} \label{app:1875}

\begin{figure}
\centering
\includegraphics[width=9cm]{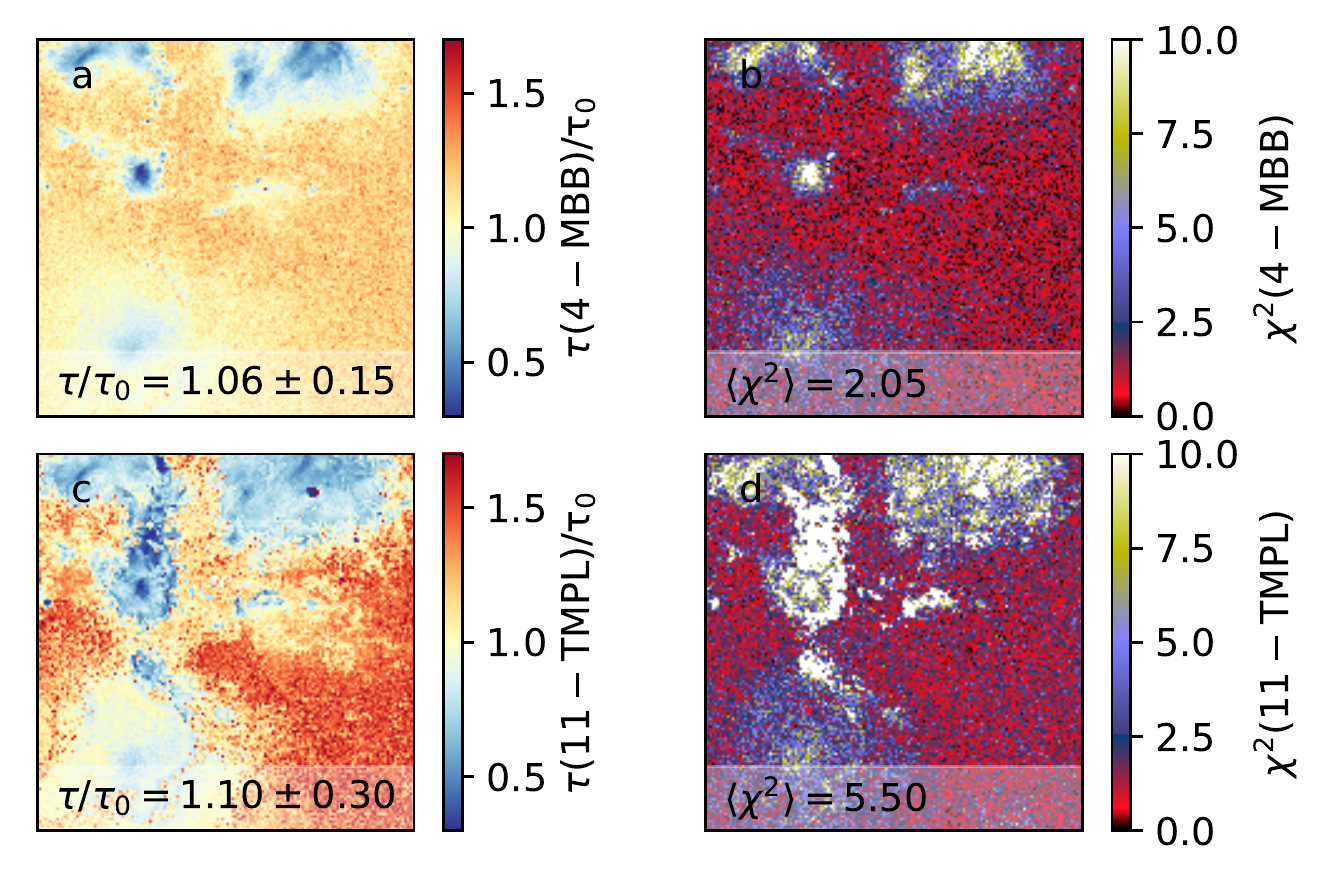}
\caption{
Fits of the 11-TMPL and 4-MBB SED models to full IRDC cloud maps
(1875$\times$1875 pixels). The parameters are the same as in
Fig.~\ref{fig:TEST_IRDC_N11}.
}
\label{fig:1875}
\end{figure}

Figure~\ref{fig:1875} shows 11-TMPL and 4-MBB fits for the whole 
1875$\times$1875 pixel maps of the IRDC cloud model. The fit parameters are
the same as in Fig.~\ref{fig:TEST_IRDC_N11} in Sect.~\ref{sect:IRDC11}, where
the smaller 256$\times$256 pixel maps correspond to the upper left corner of
these full maps.

\begin{figure*}
\sidecaption
\centering
\includegraphics[width=18cm]{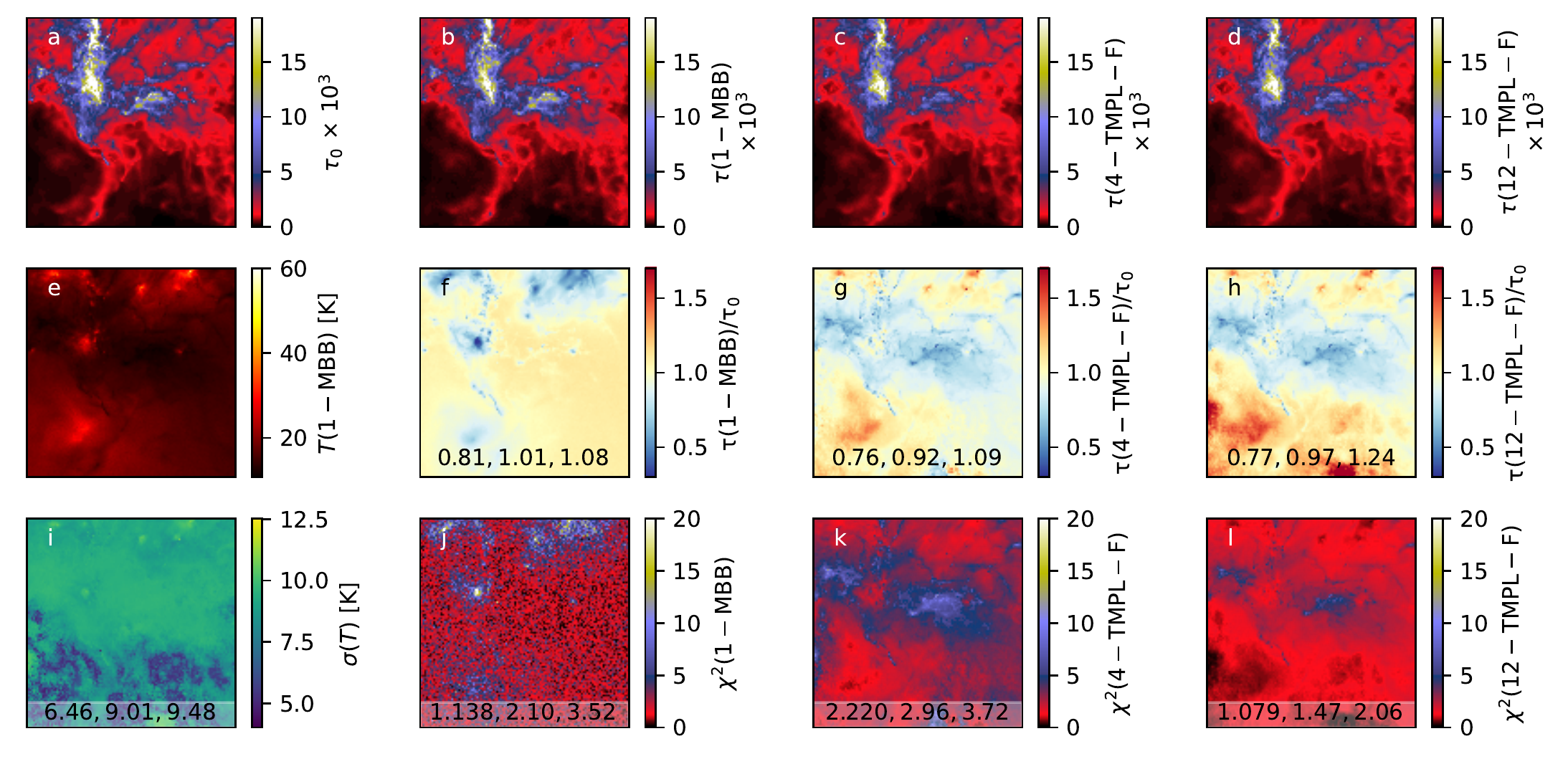}
\caption{
Comparison of 1-MBB fits and 4-TMPL-F and 12-TMPL-F fits performed in Fourier
space. The parameters of the fits are the same as in
Fig.~\ref{fig:TEST_IRDC_N11}, which corresponded to the upper left corner of
the full IRDC maps.
}
\label{fig:1875fou}
\end{figure*}

Section~\ref{sect:IRDC_FOU} discussed fits of the 4-TMPL and 12-TMPL models
with optimisation in the Fourier space. Figure~\ref{fig:1875fou} shows fits
for the full 1875$\times$1875 pixel maps using the same parameters.

\section{MCMC calculations and estimated source sizes}
\label{app:MCMC}

Although MCMC calculations are expected to be more time consuming than direct
ML calculations, they turned out to be feasible for maps of moderate size. In
the following, we show one example where the fit of the $3-TMPL$ model is done
with a basic MCMC simulation. The results are used to examine the estimated
source sizes in the case of sources with radial temperature gradients.

The synthetic observations consist of surface brightness maps at 100, 160,
250, 350, and 500\,$\mu$m, with angular resolutions of 3, 3, 4.3, and 6.1
pixels, respectively. The background has a temperature of $T_1$=15\,K and an
optical depth of one unit. We add to the maps Gaussian sources that have a
size of $FWHM$=20 pixels and a peak optical depth of two units. Each source has
a linear temperature gradient from 15\,K at the distance of $r=FWHM$ to a
lower or a higher temperature $T_2$ at the source centre. The temperature
$T_2$ of the individual sources is varied between 11\,K and 19\,K. We add to
the data white noise that corresponds to 1\% relative errors in the surface
brightness measurements.

The data were fitted with the 1-MBB model (resolution 6.1 pixels) and with the
3-TMPL model (nominal resolution 3 pixels), using components with $T_{\rm
C}$=12.5, 15.0, and 17.5\,K. No priors were used, except the requirement of
non-negative weights in the 3-TMPL fit. The MCMC calculations of the 3-TMPL
model included 200 000 MCMC steps, which was enough for reproducible results.

\begin{figure*}
\sidecaption
\centering
\includegraphics[width=12cm]{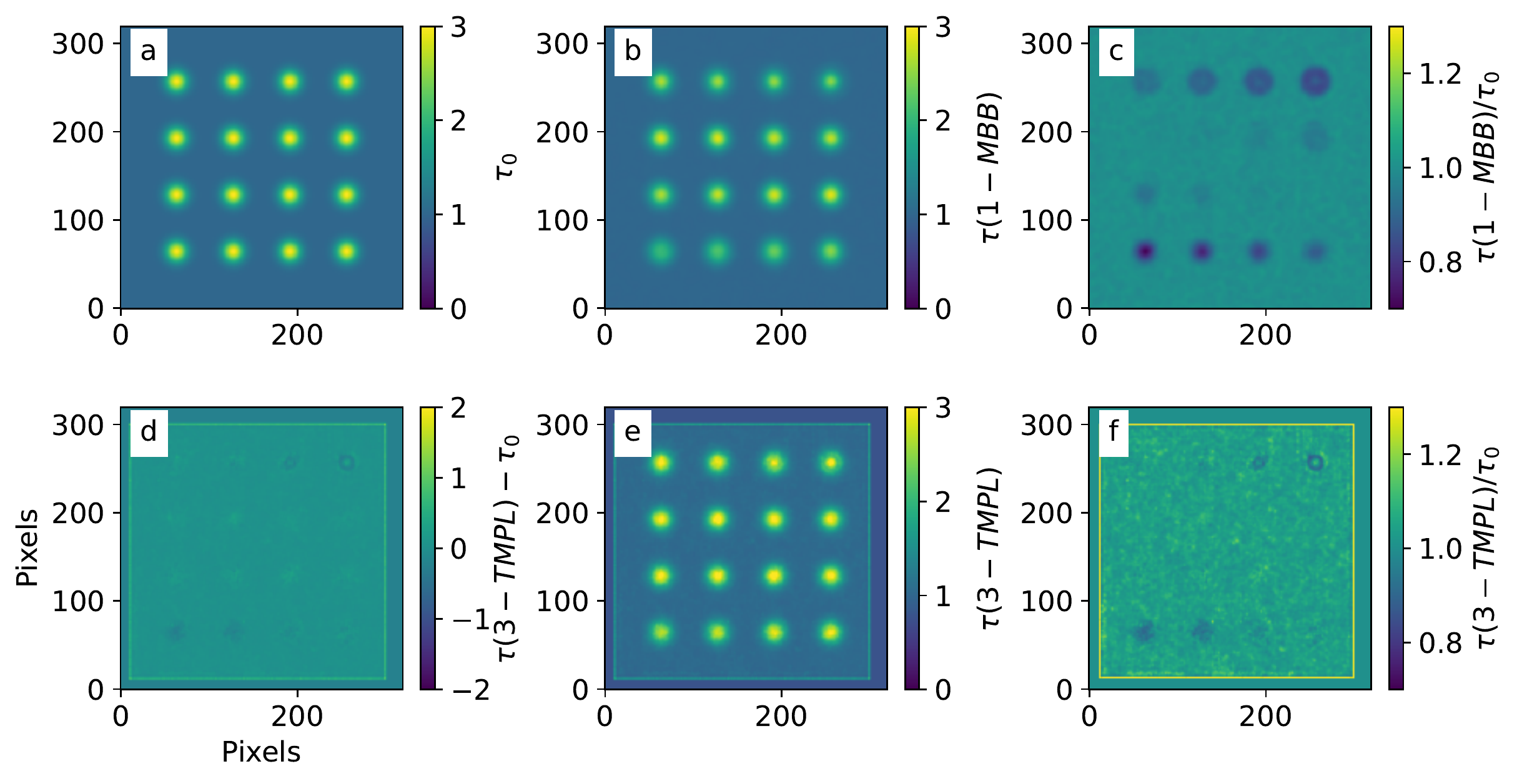}
\caption{
Fits of 1-MBB and 3-TMPL models to maps containing 16 Gaussian sources with
different radial temperature gradients. The upper frames show the true value
$\tau_0$ (frame a), the 1-MBB estimates (frame b), and the ratio of the 1-MBB
estimates divided by the true values (frame c). Data in frames b-c have a
resolution of 6.1 pixels. The lower frames show results of the 3-TMPL fit: the
fit residuals (frame d), the estimated $\tau$ values (frame e), and the ratio
of the 3-TMPL estimates and the true values (frame f). The nominal resolution
in the lower maps is equal to 3 pixels.
}
\label{fig:MCMC_map}
\end{figure*}

Figure~\ref{fig:MCMC_map} shows a simulation of $320 \times 320$ pixel maps
that contain 16 Gaussian sources. In the 1-MBB estimates, some residuals are 
visible for most sources (Fig.~\ref{fig:MCMC_map}c). The 3-TMPL fits have a
higher resolution but clear errors are visible only a couple of the coldest
and warmest sources (Fig.~\ref{fig:MCMC_map}d,f).

\begin{figure}
\centering
\includegraphics[width=8.8cm]{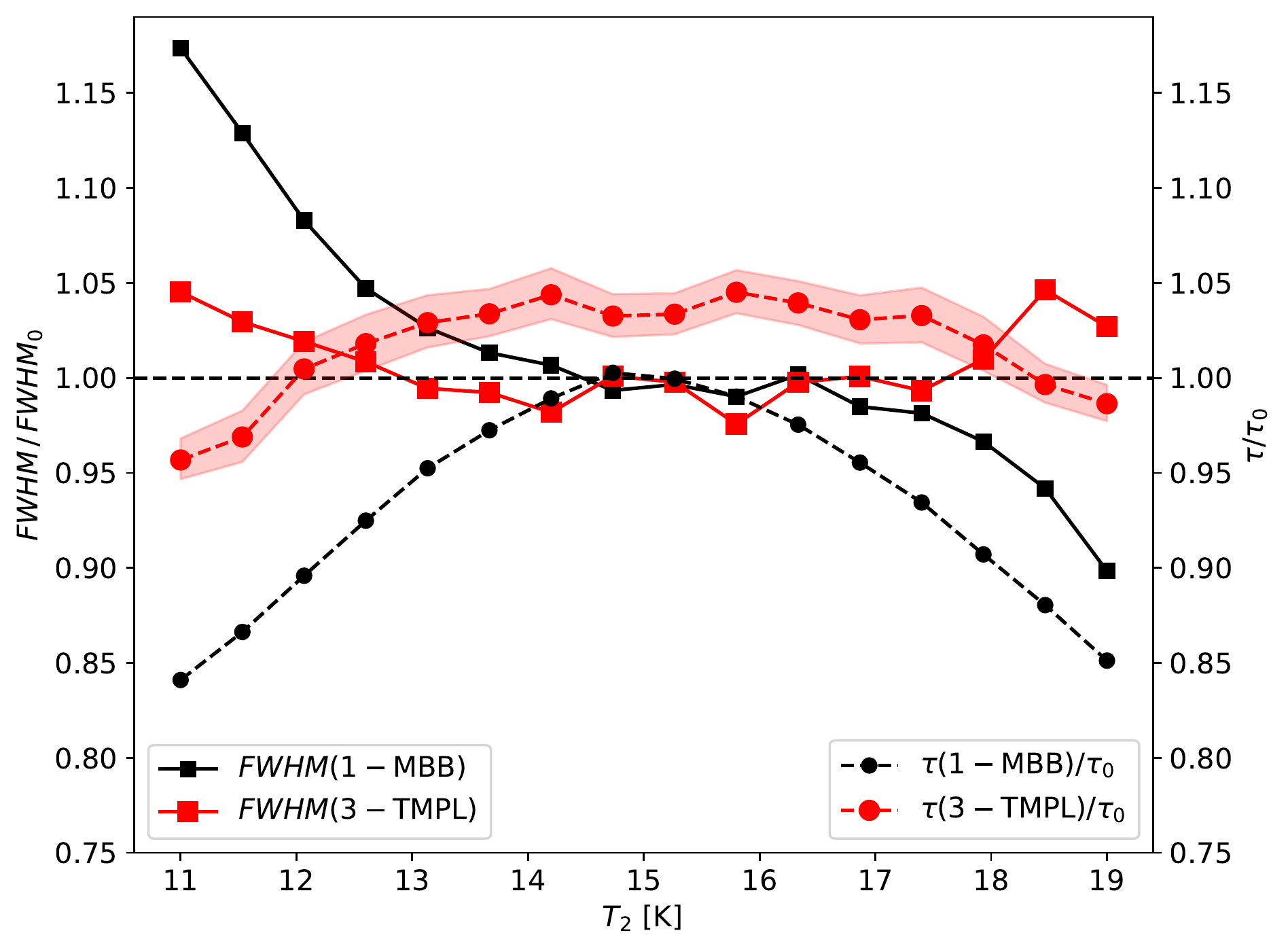}
\caption{
Test with sources with radial temperature gradients. The average optical
depth estimates (circles) and the source FWHM estimates (squares) are plotted
relative to their correct values. The symbols correspond to 16 sources that
are plotted against their central temperature $T_2$. The 1-MBB and the 3-TMPL
results are plotted in black and red, respectively. The shading shows the MCMC
error estimates for the $\tau/\tau_0$ ratio in the 3-TMPL fits.
}
\label{fig:MCMC_plot}
\end{figure}

We fitted each of the 16 sources with Gaussians to study the errors in the
estimated source sizes. To make the results comparable between the two fits,
the 3-TMPL predictions were first convolved down to the resolution of the
1-MBB map.  Figure~\ref{fig:MCMC_plot} shows the ratios of the estimated and
expected beam-convolved FWHM sizes for the 16 sources in
Fig.~\ref{fig:MCMC_map}. The x-axis corresponds to the temperature $T_2$ at
the centre of each source.  The 1-MBB fits are seen to overestimate the size
of the coldest source by $\sim$15\% and underestimate the size of the warmest
source by 10\%. In the 3-TMPL fit the errors remain below $\sim$5\%.

We calculated for each source also mean optical depths over and area of
40$\times$40 pixel (twice the source FWHM). Unsurprisingly, the bias of the
1-MBB estimates increases with the temperature difference $|T_1-T_2|$, and the
optical depth is underestimated by up to 15\% for both the coldest and warmest
sources. The 3-TMPL estimates show a positive bias of $\sim$4\%, with values
dropping when $T_2$ goes outside the range of $T_{\rm C}$ values.  Because of
noise (and requirement of non-negative component weights) the 3-TMPL model is
thus generally slightly overestimating the width of the temperature
distribution. The results are on average more accurate than for the 1-MBB
model. This is partly due to a better description of the temperature
variations and partly due to the higher effective resolution, which reduces
the mixing of temperatures within the beam. 

Figure~\ref{fig:MCMC_plot} also shows the MCMC error estimates for the optical
depths. The shaded region corresponds to the interquartile range in the MCMC
samples over the 40$\times$40 pixels area around each source. In the plot, the
raw error estimates (per resolution element) are scaled down by the ratio
between 40 pixels (the averaged area) and nominal resolution of 3 pixels. The
MCMC error estimates seem to be consistent with the scatter between the
sources (of similar $T_2$) but of course does not cover the larger component
of systematic errors.

For the above 3-TMPL fit of the 320$\times$320 pixel map, the MCMC run took
about ten minutes and is thus not significantly slower than the ML solution.
In our tests, the corresponding analysis of 576$\times$576 pixel maps took
about half an hour, in agreement with linear dependence on the number of
pixels. However, all calculations were done using a GPU, and the run time on a
CPU would be at least one order of magnitude longer.

\section{MCMC calculations with an alternative SED model}  \label{app:MCMC2}

In this paper the $N$-TMPL and $N$-MBB models were based on a set of a few
discrete temperatures. However, each template spectrum of the $N$-TMPL models
and even the basis functions of the $N$-MBB models could equally be integrals
over some temperature distributions.

As a basic example, we tested a three-parameter model where the parameters are
the optical depth, the central temperature $T_{\rm C}$ and the width $\delta
T$ of a Gaussian temperature distribution. A grid of SED templates was first
calculated as the function of $T_{\rm C}$ and $\delta T$. The fitted
parameters were the indices to the SED grid and the scaling of the selected
template spectrum. As in Sect.~\ref{app:MCMC}, the model has three free
parameters and the fit was performed with MCMC methods. We refer to this as
the GRID model.

\begin{figure}
\centering
\includegraphics[width=8.8cm]{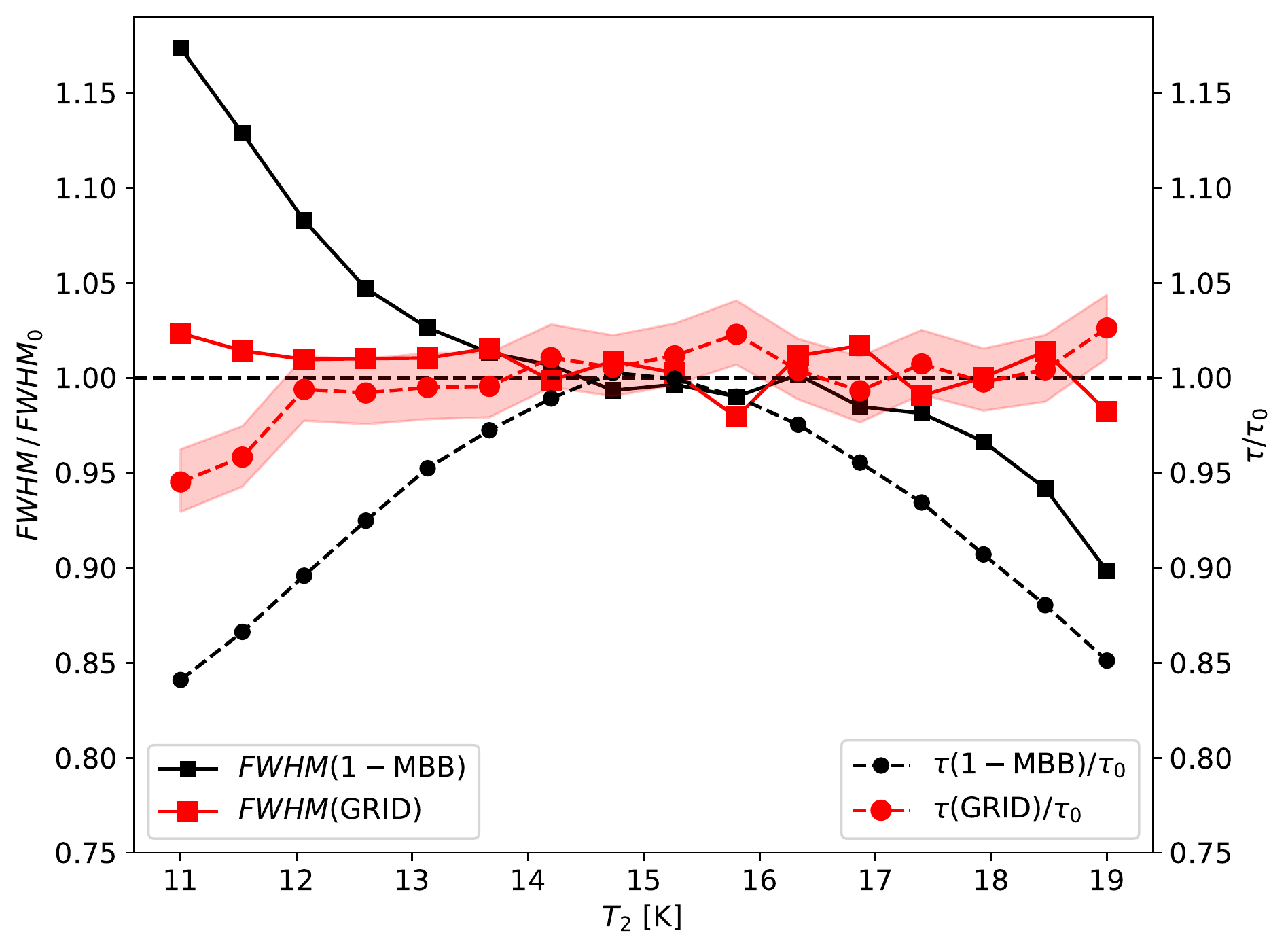}
\caption{
Alternative fit of sources with radial temperature gradients. The symbols
are as in Fig.~\ref{fig:MCMC_plot}, but the red colour and the shaded error
regions are for the GRID model.
}
\label{fig:MCMC2_plot}
\end{figure}

Figure~\ref{fig:MCMC2_plot} shows the results for the same synthetic
observations as in Sect.~\ref{app:MCMC} (maps of 320$\times$320 pixels).
Because these consist of two discrete temperatures (with spatially varying
$T_2$), the fitted model of a single Gaussian temperature distribution does
not exactly match the simulated data. The results are improved compared to the
3-TMPL fits in Fig.~\ref{fig:MCMC_plot}, and the $\tau$ values show now
very little bias. On the other hand, the run time (on GPUs) is also longer, as
the calculation of three MBB functions is replaced by the more
time-consuming look-up from a SED table. After an initial burn-in phase, the
calculation of 200 000 MCMC steps took one hour.

\begin{figure}
\centering
\includegraphics[width=8.8cm]{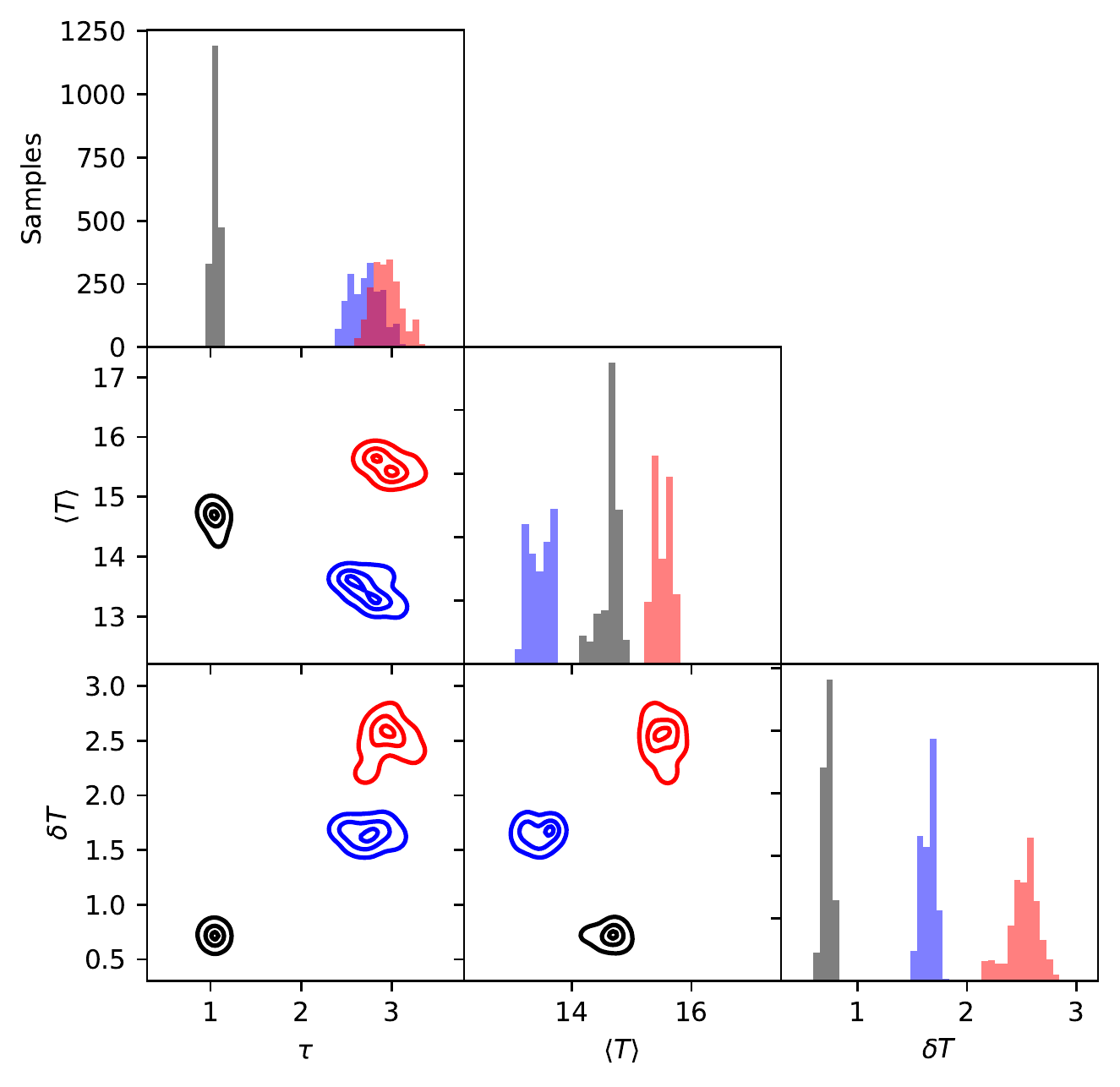}
\caption{
Corner plot showing parameter correlations for three areas in the fits of 
Fig.~\ref{fig:MCMC2_plot}. The parameters are the optical depth $\tau$, the
mean temperature $T_{\rm C}$, and the standard deviation of the temperature
distribution $\delta T_{\rm C}$. The data correspond to averages over $3\times
3$ pixels towards the centre of the coldest source (blue), a position in the
15\,K background (black), and the centre of the warmest source (red colour).
The contours are drawn at 10\%, 50\%, and 90\% of the peak probability
density.
}
\label{fig:MCMC2_corner}
\end{figure}

MCMC calculations provide complete information of the posterior probability of
the fitted parameters: the optical depth $\tau$, the mean temperature $\langle
T \rangle$, and the width of the temperature distribution $\delta T$.
Figure~\ref{fig:MCMC2_corner} shows examples of these distributions. With the
high S/N of the observations assumed in this example, there are no
strong correlations between the model parameters.

\begin{figure}
\centering
\includegraphics[width=8.8cm]{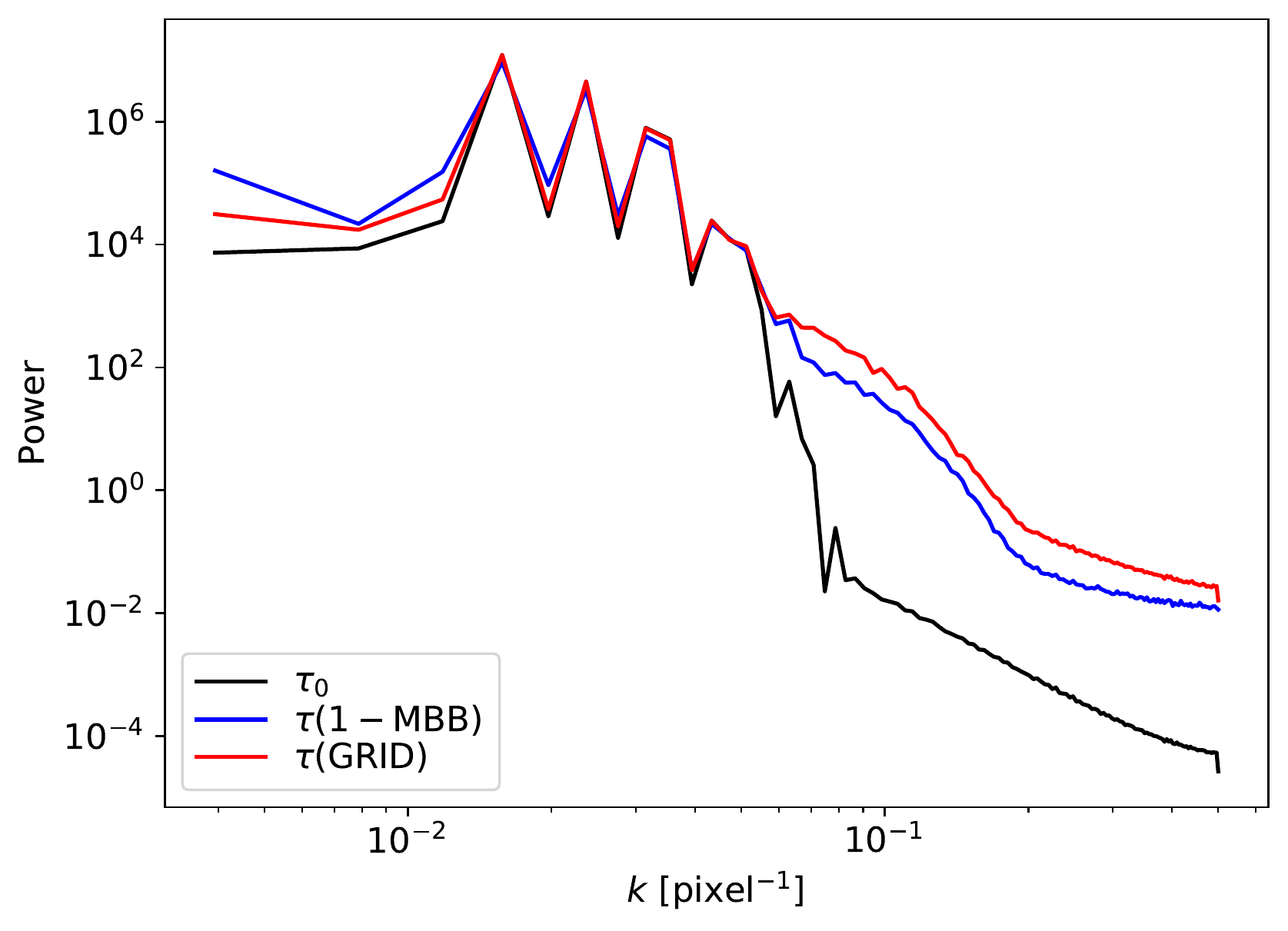}
\caption{
Power spectra of the optical depth maps for the fits in
Fig.~\ref{fig:MCMC2_plot}. The spectra are shown for the true optical depth
$\tau_0$ and the 1-MBB and GRID estimates. For the comparison, the other maps
are convolved to the resolution of the 1-MBB map.
}
\label{fig:ps}
\end{figure}

Figure~\ref{fig:ps} compares the power spectra of the different optical depth
maps. The analysed region excludes the borders where $\tau$(GRID) solution has
not been calculated (cf. Fig.~\ref{fig:MCMC_map}f). When the GRID map is
convolved to the resolution of the 1-MBB map, the power spectra are nearly
identical up to $k\sim 0.05 \, {\rm pixel}^{-1}$. We assumed here that, before
this convolution, the GRID model would have the nominal resolution of 3 pixels
compared to the 6.1 pixel resolution of the 1-MBB model. Figure~\ref{fig:ps}
is approximately consistent with this assumption. The 1-MBB and GRID power
spectra separate only at high spatial frequencies, presumably because the GRID
solution has a higher statistical noise, even when degraded to the resolution
of the 1-MBB solution.

\end{document}